\documentclass[11pt]{article} \hoffset=-15mm \voffset=-10mm
\usepackage{graphicx,amsmath,amssymb,epsf} \usepackage{latexsym,bm,
  slashed, cite} \usepackage{xcolor} \definecolor{dark}{rgb}{0.10,0.2,0.3}
\definecolor{magenta}{rgb}{0.7,0.1,0.3}
\definecolor{purpure}{rgb}{0.5,0.15,0.3}
\usepackage[font=small,format=plain,labelfont=bf,up,textfont=it,up]{caption}
\usepackage{hyperref, cite} \hypersetup{colorlinks, citecolor=blue,
  filecolor=blue, linkcolor=magenta,
  urlcolor=purpure,hyperfootnotes=true,pdftex} 
  
\newcommand{\tr}{{\rm tr}}
\newcommand{\bk}{{\bm k}}
\newcommand{\bp}{{\bm p}}
\newcommand{\bq}{{\bm q}}
\newcommand{\br}{{\bm r}}
\newcommand{\CA}{C_{\hspace{-0.3ex}A}}
\newcommand{\CF}{C_{\hspace{-0.1ex}F}}
\newcommand{\KaTie}{\mbox{\sc Ka\hspace{-0.2ex}Tie}}
\newcommand{\hatC}{\hat{C}}
\newcommand{\Equation}[1]{Eq.~(\ref{#1})}
\newcommand{\PlusDist}[1]{\left[\frac{#1}{\br^2}\right]_+}

\usepackage[makeroom]{cancel}
\usepackage[normalem]{ulem}

\definecolor{kkcolor}{rgb}{1,0,0}

\newcommand\kkout{\marginpar{\color{kkcolor}$\clubsuit$}\bgroup\markoverwith{\color{kkcolor}{\rule[04ex]{2pt}{0.8pt}}}\ULon}

\definecolor{avhcolor}{rgb}{1,0,0}

\newcommand\avhout{\marginpar{\color{avhcolor}$\clubsuit$}\bgroup\markoverwith{\color{avhcolor}{\rule[04ex]{2pt}{0.8pt}}}\ULon}

 \title{\bf  \Large 
Forward Higgs production within high energy factorization  in the heavy quark limit    at next-to-leading order accuracy   
   } \author{M.~Hentschinski{}$^a$, K.~Kutak{}$^b$, A.~van~Hameren{}$^b$ \\ \\
{}$^a$ Departamento de Actuaria, F\'isica y Matem\'aticas, \\
Universidad de las Americas Puebla, \\ Santa Catarina Martir, 72820 Puebla, Mexico \\[1ex]
{}$^b$  Institute of Nuclear Physics, Polish Academy of Sciences, \\ 
 ul.~Radzikowskiego 152, 31-342, Krak\'ow, Poland\\
}

\begin{document}

\maketitle
\begin{abstract}
  We use Lipatov's high energy effective action to determine the
  next-to-leading order corrections to Higgs production in the forward
  region within high energy factorization making use of the infinite
  top mass limit.  Our result is based on an explicit calculation of
  real corrections combined with virtual corrections determined
  earlier by Nefedov. As a new element we provide a proper definition
  of the desired next-to-leading order coefficient within the high
  energy effective action framework, extending a previously proposed
  prescription.  We further propose a subtraction mechanism to achieve
  for this coefficient a stable cancellation of real and virtual
  infra-red singularities in the presence of external off-shell
  legs. Apart from its relevance for direct phenomenological studies,
  such as high energy resummation of Higgs $+$ jet configurations, our
  result will be further of use for the study of transverse momentum
  dependent factorization in the high energy limit.
 \end{abstract}

\section{Introduction}
\label{sec:intro}

The observation of the Higgs boson by the ATLAS and CMS experiments
\cite{Aad:2012tfa,Chatrchyan:2012ufa} confirmed expectations that the
Standard Model of particle physics is a consistent theory of strong
and electro-weak interactions.  The electro-weak sector of the
Standard Model is being explored now in detail by measuring properties
of the Higgs boson \cite{DiMicco:2019ngk}.  The success of its
discovery is complemented with an advancement in 
techniques for the calculation of production cross sections and decay rates with high accuracy; for a recent review see
\cite{Heinrich:2020ybq}. To determine the cross section for
Higgs production in the central rapidity region, which is the dominant
production region,  one usually uses the framework of collinear factorization, where the incoming partons are collinear with the beam axis
and are approximately on shell.  In this framework one can reliably
calculate the production cross section up to next-to-next-to-leading
order (NNLO) accuracy~\cite{Dreyer:2020xaj}, complemented with Monte
Carlo simulations to describe complete production events including
hadronization and detector simulation \cite{Monni:2020nks}.\\

In this paper we use on other hand the process of Higgs boson production in the forward direction  to advance further the formulation of  high energy factorization \cite{Catani:1990eg,Collins:1991ty} for Quantum Chromodynamics (QCD) to next-to-leading order (NLO)
accuracy. To be precise, our discussion is based on Lipatov's high energy effective action \cite{Lipatov:1995pn,Lipatov:1996ts} and the determination of NLO correction will be achieved within this framework. From a phenomenolgical point of view, for center of mass energies accessible at the Large Hadron Colliders, cross-sections for forward production of Higgs bosons are most likely too small to be observed. Our result is therefore at first of formal interest and serves to explore further the proper definition of NLO coefficients within high energy factorization. We nevertheless would like to stress that there exist already studies which investigate the relevance of high energy resummation for Higgs-jet configurations \cite{Celiberto:2020tmb}. We certainly expect that our result will be of relevance for the phenomenology of future colliders such as the Future Circular Collider project \cite{Mangano:2017tke}. \\

The formalism of high energy factorization was developed in
order to resum perturbative contributions to the cross section,
enhanced by logarithms in the center-of-mass energy, which are of
relevance whenever the center-of-mass energy of the process is much larger
than any other scale involved \cite{Fadin:1975cb, Lipatov:1976zz,
  Kuraev:1977fs, Balitsky:1978ic}. This resummation is also applicable
when the configuration of the final state is such that at least one of
the final state particle is produced in the forward direction, where
the logarithmic dependence on the center-of-mass energy translates
into large differences in rapidity.  The resulting framework is then
known under the term hybrid factorization \cite{Dumitru:2005gt}, see
also \cite{Marquet:2007vb,Deak:2009xt,Chachamis:2015ona}.\\

With production taking place in the
forward region of one of the hadrons, partons originating from this
hadron are characterized by relatively large momentum fractions $x$
with the corresponding parton distribution subject to conventional
DGLAP evolution. Partons stemming from the second hadron are on the
other hand characterized by a very small longitudinal momentum
fraction $x$ as well as a non-zero transverse momentum $k_T$. With
quark exchange power-suppressed in the vacuum channel, the
corresponding unintegrated gluon distribution is subject to the
Balitsky-Fadin-Kuraev-Lipatov (BFKL) equation
\cite{Fadin:1975cb, Lipatov:1976zz,
  Kuraev:1977fs, Balitsky:1978ic,
  Fadin:1998py,Ciafaloni:1998gs} and its nonlinear extensions
\cite{Kovchegov:1999yj,Balitsky:1995ub,JalilianMarian:1997jx,JalilianMarian:1997gr,Kovner:2000pt,Kovner:1999bj}
respectively.\\

The high energy factorization formalism was rather successful in
describing production of final states widely separated in rapidity
\cite{Ducloue:2013hia, Caporale:2016zkc,
  Celiberto:2016ygs,Chachamis:2017vfa, Caporale:2018qnm} and processes
in the forward rapidity region
\cite{vanHameren:2014ala,vanHameren:2015uia, Bautista:2016xnp,
  Celiberto:2018muu, Garcia:2019tne} and some processes are even known
at NLO accuracy both in momentum space \cite{Bartels:2001ge,
  Bartels:2002yj,
  Hentschinski:2014esa,Hentschinski:2014bra,Hentschinski:2014lma,Chachamis:2012cc,Celiberto:2017ptm}
and coordinate space
\cite{Boussarie:2019ero,Boussarie:2016ogo,Beuf:2017bpd}. It is worthwhile to note that there has
been progress during the recent years in the development of
computational methods which allowed for a reformulation of the
evaluation of high energy factorization matrix elements
\cite{vanHameren:2012uj,vanHameren:2012if,vanHameren:2013csa} as well as the
subsequent automation of tree level matrix elements and the
calculation of parton-level cross sections via Monte Carlo methods
\cite{vanHameren:2016kkz}.\\

A very useful tool to calculate the $k_T$ dependent
matrix elements, which arise from the aforementioned factorization procedure, is then provided  by the previously mentioned  high energy effective action
\cite{Lipatov:1995pn,Lipatov:1996ts}. It yields  matrix elements
for the interaction between conventional QCD fields and reggeized
gluon fields, localized in rapidity. The reggeized gluon field is in
this context an auxiliary degree of freedom which is used to formulate
gauge invariant factorization of QCD amplitudes, see
\cite{Hentschinski:2020rfx} for a recent review and
\cite{Bock:2020cnd} for the determination of a corresponding effective
action for electro-weak fields. With the
present project we plan to revisit the strategy of calculations of NLO
processes using Lipatov's high energy effective action, through
considering a colorless massive final state, see
\cite{Hautmann:2002tu,Lipatov:2005at,Lipatov:2009qx,Lipatov:2015vya}
for the tree level result within $k_T$ factorization. The simplicity
of the final state allows to address in a well structured manner all
aspects of a complete NLO calculations and to give a compact formula for
the cross section which can be presented analytically and eventually
implemented in a numerical code, where the latter is left as a task for the future. As we will show in the following, the high energy effective action provides a well defined and manifestly gauge invariant setup for such calculations. 
\\

The paper is organized as follows. In Sec.~\ref{sec:eff} we provide a quick overview over the high energy factorization as formulated within the high energy effective actio and how it can be used for the actual calculation of NLO corrections. In Sec.~\ref{sec:partonic} we then present the results of our calculations, while in Sec.~\ref{sec:concl} we draw our conclusions. Some details of our calculations have been referred to the Appendix \ref{sec:finiteness}.

\section{The High-Energy Effective Action}
\label{sec:eff}

Our calculation is based on  Lipatov's high energy effective action
\cite{Lipatov:1995pn}. Within this framework, QCD amplitudes are in
the high energy limit decomposed into gauge invariant sub-amplitudes
which are localized in rapidity space and describe the coupling of
quarks ($\psi$), gluon ($v_\mu$) and ghost ($\phi$) fields to a new
degree of freedom, the reggeized gluon field $A_\pm (x)$. The latter
is introduced as a convenient tool to reconstruct the complete QCD
amplitudes in the high energy limit out of the sub-amplitudes
restricted to small rapidity intervals. To be explicit we consider scattering of two partons with momenta $p_a$ and $p_b$ which serve to define the light-cone directions of the high energy effective action
\begin{align}
  \label{eq:nplusminus}
  (n^\pm)^\mu & = \frac{2}{\sqrt{\hat{s}}} p_{a,b}^\mu, & \hat{s} & = 2 p_a \cdot p_b ,
\end{align}
which yields the following Sudakov decomposition of a generic four-momentum, 
\begin{align}
k & = k^+ \frac{n^-}{2} +  k^- \frac{n^+}{2} + k_T, & k^\pm &= k \cdot n^\pm, & n^\pm \cdot k_T & = 0
~.
\end{align}
Here, $k_T$ is the embedding of the Euclidean vector $\bk$ into Minkowski space, so $k_T^2=-\bk^2$.
Lipatov's effective action is then obtained by adding an induced term $
S_{\text{ind.}}$ to the QCD action $S_{\text{QCD}}$,
\begin{align}
  \label{eq:effac}
S_{\text{eff}}& = S_{\text{QCD}} +
S_{\text{ind.}}\; ,
\end{align}
where the induced term $ S_{\text{ind.}}$ describes the coupling of
the gluonic field $v_\mu = -it^a v_\mu^a(x)$ to the reggeized gluon
field $A_\pm(x) = - i t^a A_\pm^a (x)$.  High energy factorized
amplitudes reveal strong ordering in plus and minus components of
momenta which is reflected in the following kinematic constraint
obeyed by the reggeized gluon field:
\begin{align}
  \label{eq:kinematic}
  \partial_+ A_- (x)& = 0 = \partial_- A_+(x).
\end{align}
Even though the reggeized gluon field is charged under the QCD gauge
group SU$(N_c)$, it is invariant under local gauge transformation
$\delta A_\pm = 0$.  Its kinetic term and the gauge invariant coupling
to the QCD gluon field are contained in the induced term
\begin{align}
\label{eq:1efflagrangian}
  S_{\text{ind.}} = \int \text{d}^4 x \,
\text{tr}\left[\left(W_-[v(x)] - A_-(x) \right)\partial^2_\perp A_+(x)\right]
+\text{tr}\left[\left(W_+[v(x)] - A_+(x) \right)\partial^2_\perp A_-(x)\right],
\end{align}
with 
\begin{align}
  \label{eq:funct_expand}
  W_\pm[v(x)] =&
v_\pm(x) \frac{1}{ D_\pm}\partial_\pm,
&
D_\pm & = \partial_\pm + g v_\pm (x).
\end{align}

\begin{figure}[htb]
    \label{fig:subfigures}
   \centering
   \parbox{.7cm}{\includegraphics[height = 1.8cm]{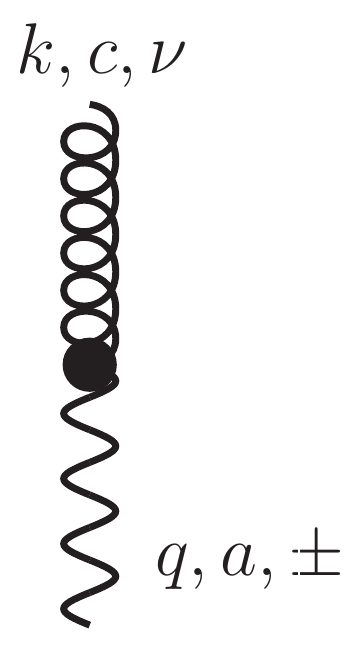}} $=  \displaystyle 
   \begin{array}[h]{ll}
    \\  \\ - i{\bm q}^2 \delta^{a c} (n^\pm)^\nu,  \\ \\  \qquad   k^\pm = 0.
   \end{array}  $ 
 \parbox{1.2cm}{ \includegraphics[height = 1.8cm]{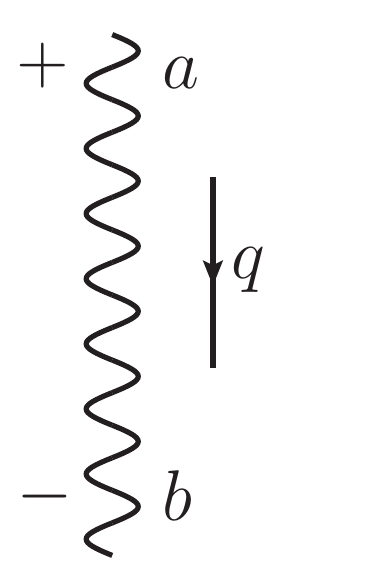}}  $=  \displaystyle    \begin{array}[h]{ll}
    \delta^{ab} \frac{ i/2}{{\bm q}^2} \end{array}$ 
 \parbox{1.7cm}{\includegraphics[height = 1.8cm]{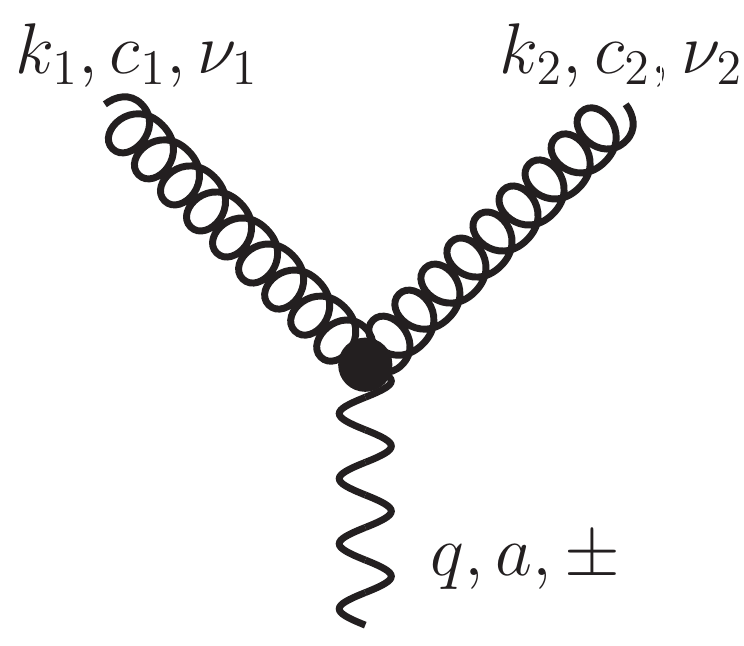}} $ \displaystyle  =  \begin{array}[h]{ll}  \\ \\ g f^{c_1 c_2 a} \frac{{\bm q}^2}{k_1^\pm}   (n^\pm)^{\nu_1} (n^\pm)^{\nu_2},  \\ \\ \quad  k_1^\pm  + k_2^\pm  = 0
 \end{array}$
 \\
\parbox{4cm}{\center (a)} \parbox{4cm}{\center (b)} \parbox{4cm}{\center (c)}

 \caption{\small Feynman rules for the lowest-order effective vertices of the effective action. Wavy lines denote reggeized fields and curly lines gluons. }
\label{fig:3}
\end{figure}

For a more in depth discussion of the effective action we refer to the
 reviews \cite{Chachamis:2012mw}. Due to the induced term in
Eq.~(\ref{eq:effac}), the Feynman rules of the effective action
comprise, apart from the usual QCD Feynman rules, the propagator of
the reggeized gluon and an infinite number of so-called induced
vertices.  Vertices and propagators needed for the current study are
collected in Fig.~\ref{fig:3}.
Determination of NLO corrections using this effective action approach
has been addressed recently  to a certain extent, through the explicit
calculation of the NLO corrections to both quark
\cite{Hentschinski:2011tz} and gluon \cite{Chachamis:2012cc} induced
forward jets (with associated radiation) as well as the determination
of the gluon Regge trajectory up to 2-loops
\cite{Chachamis:2012gh,Chachamis:2013hma}. These previous applications
have all in common that they are, at amplitude level, restricted to a
color octet projection and, therefore, single reggeized gluon
exchange. Due to the particular color structure of the reggeized gluon
field, which is restricted to the anti-symmetric color octet, see
Fig.~\ref{fig:3} and \cite{Lipatov:1995pn, Hentschinski:2011xg},
color singlet exchange requires to go beyond a single reggeized gluon
exchange and to consider the two reggeized gluon exchange
contribution. For a discussion of the analogous high energy effective
for flavor exchange \cite{Lipatov:2000se} at NLO see {\it e.g.}
\cite{Nefedov:2017qzc,Nefedov:2018vyt,Nefedov:2019mrg}. 

\subsection{Factorization of partonic cross sections in the high energy limit}
\label{sec:frameworkNLO}
In the following we describe the framework to be used to determine the NLO corrections to the forward Higgs impact factor. The formulation of this framework is based on the explicit results obtained in the case of forward quark and gluon jets \cite{Hentschinski:2011tz, Chachamis:2012cc}  as well as the determination of the gluon Regge trajectory up to 2-loops \cite{Chachamis:2012gh,Chachamis:2013hma}, where the later only addresses  virtual corrections. 
Adapting a normalization of impact factors motivated by $k_T$-factorization, we factorize in the high energy limit the partonic cross section into
\begin{align}
  \label{eq:dsigKT}
  d \hat{\sigma}_{ab} & = \hat{h}_a^{k_T}(\bk) \hat{h}_b^{ugd}(\bk)  \frac{d^{2 + 2 \epsilon} \bk}{\pi^{1 + \epsilon}},
\end{align}
where we regulate both infra-red and ultra-violet corrections using dimensional regularization in $d = 4 + 2 \epsilon$ dimensions.
Here $ \hat{h}_a^{k_T}$ denotes the impact parameter in the fragmentation region of parton $a$ while $\hat{h}_b^{ugd}$ is the impact factor\footnote{The subscript `ugd' refers to unintegrated gluon density. It  denotes  that the normalization of this impact factor is in accordance with  an unintegrated gluon density at partonic level} in the fragmentation of parton $b$. While the normalization is adapted to the asymmetric scenario where the transverse scale in the fragmentation region of parton $a$ is significantly larger than the corresponding scale in the fragmentation region of parton $b$, our framework is completely general and does not assume a priori such a hierarchy. In terms of (off-shell) matrix elements of reggeized gluon fields $r^\pm$ and conventional QCD fields we have 
\begin{align}
  \label{eq:impact_factors}
  \hat{h}_{b}^{ugd}(\bk) & = \frac{(\pi)^{d/2}}{2 p_b^-} \int dk^+ 
\frac{\overline{|\mathcal{M}_{br^- \to X_b^n}|^2}}{ \bk^2}
d \Phi^{(n)}  \delta^{(d)}(p_b + k - \sum_{j=1}^n p_j ), \notag \\
 \hat{h}^{k_T}(\bk) & =    \int \frac{dk^-}{k^-} d\hat{\sigma}_{a+} ,
\end{align}
where off-shell partonic cross section and corresponding off-shell squared matrix elements $|\mathcal{A}|^2$ are in terms of effective action matrix elements  obtained as:
\begin{align}
  \label{eq:dsig_hat}
  d\hat{\sigma}_{a+} & = \frac{1}{2 p_a^+ k^-} \frac{ \overline{|\mathcal{A}_{ar^+  \to   X_a^{(n)}}|}^2}{N_c^2-1} (2 \pi)^d \delta^d(p_a + k - \sum_{j=1}^n p_j) d \Phi^{(n)} \notag \\
&  \overline{|\mathcal{A}_{ar^+  \to   X_a^{(n)}}|^2}
=
 \frac{(k^-)^2}{4 \bk^2}\overline{|\mathcal{M}_{ar^+  \to   X_a^{(n)}}|^2};
\end{align}
$X_{a,b}^{(n)}$ denotes any $n$-particle system produced in the regarding fragmentation region,
 \begin{align}
  \label{eq:PS}
 d \Phi^{(n)} & = \prod_{j=1}^n \frac{d^d p_j}{(2 \pi)^{d-1}}  \delta_+(p_j^2 - m_j^2),
\end{align}
 the $n$-particle phase space and we average and sum over spin and color of incoming and produced particles respectively. Apart from production in the fragmentation region, there exists also the possibility of production at central rapidities. Within high energy factorization as provided by the  high energy effective and restricting to processes with only one reggeized gluon exchange, this is described through the collision of two reggeized gluons with opposite polarizations. We have
\begin{align}
  \label{eq:dsig_pieces}
  d \hat{\sigma}_{ab}& =  \hat{h}_a^{(kT)}(\bk_1)\hat{h}_b^{(ugd)}(-\bk_2) V(-\bk_1, \bk_2)
\frac{d^{2 + 2 \epsilon} \bk_1}{\pi^{1 + \epsilon}} d^{2 + 2 \epsilon} \bk_2 d\eta,
\end{align}
with \cite{Hentschinski:2011tz}
\begin{align}
  \label{eq:prod_vertex}
  V(\bk_1, \bk_2) & =
\frac{N_c^2-1}{8} \int dM^2 \frac{\overline{|\mathcal{M}_{r^-(k_1)r^+(k_2) \to X^{(n)}_c)}|^2}}{\bk_1^2 \bk_2^2} d \Phi^{(n)} \delta^{(d)}(k_1 + k_2 - \sum_{j=1}^n p_j ),
\end{align}
where $M^2 = k_1^+ k_2^-$;   
to leading order in the strong coupling constant  one finds
\begin{align}
  \label{eq:V1}
   V(\bk_1, \bk_2) & = \frac{\alpha_s \CA}{\pi_\epsilon \pi (\bk_1 + \bk_2)^2}, & \pi_\epsilon& \equiv \pi^{1 + \epsilon} \Gamma(1 - \epsilon)\mu^{2 \epsilon}, & \alpha_s & = \frac{g^2 \Gamma(1 - \epsilon) \mu^{2 \epsilon}}{(4 \pi)^{1 + \epsilon}}.
\end{align}
For the generic inclusive process in which we are interested in, we integrate over the entire phase space of the centrally produced gluon and find that 
the integral over  rapidity $\eta = \frac{1}{2} \ln(k_1^+/k_2^-)$ in Eq.~\eqref{eq:dsig_pieces}  requires an appropriate regularization. The generic choice is  $\rho/2 > \eta > - \rho/2$ with $\rho \to \infty$. Apart from the central rapidity production vertex, there exists also virtual corrections at central rapidities. They are obtained as self-energy corrections to the reggeized gluon fields and yield the following one-loop reggeized gluon propagator \cite{Chachamis:2013hma},
\begin{align}
  \label{eq:barepropR}
   G \left(\rho; \epsilon, {\bm k}^2, \mu^2   \right)
&=
\frac{i/2}{{\bm k}^2} \left\{ 1 + \frac{i/2}{{\bm k}^2} \Sigma \left(\rho; \epsilon, \frac{{\bm k}^2}{\mu^2}    \right)   + \ldots   \right\}.
\end{align}
At the level of a partonic cross section this yields
\begin{align}
  \label{eq:dsig_ab_virt}
  d\hat{\sigma}^c_{ab} & = \hat{h}_a^{(kT)} (\bk)\hat{h}_b^{(ugd)}(\bk) 
\left| 1 + \frac{i/2}{{\bm k}^2} \Sigma \left(\rho; \epsilon, \frac{{\bm k}^2}{\mu^2}    \right)  + \ldots   \right|^2
\frac{d^{2 + 2 \epsilon} \bk}{\pi^{1 + \epsilon}}
~.
\end{align}
$\Sigma$ has a perturbative expansion in $\alpha_s$, and the first term, including this expansion parameter, is given by~\cite{Chachamis:2012cc,Chachamis:2013hma,Chachamis:2012gh}
\begin{align}
\label{eq:self_1loopRES}
   \frac{\Sigma^{(1)}\left(\rho; \epsilon, \frac{{\bm k}^2}{\mu^2}    \right) }{(-2i {\bm k}^2) }     
& = \frac{\alpha_s}{4 \pi} \left(\frac{\bk^2}{\mu^2} \right)^\epsilon\left[\frac{-\CA (2 \rho - i \pi)}{\epsilon} -  \frac{5 \CA - 2 n_f}{3 \epsilon}+ \frac{31 \CA}{9} - \frac{10 n_f}{9} \right] + \mathcal{O}(\epsilon).
\end{align}
Like the inclusive central production vertex, the virtual corrections contain a rapidity divergence which we regulate by tilting the light-cone directions of the high energy effective action against the light-cone
\begin{align}
  \label{eq:npmtilt}
  n^\pm &\to n^\pm + e^{- \rho} n^\mp & \rho & \to  \infty.
\end{align}

\subsection{Subtraction and Transition Function}
\label{sec:subtrans}

Beyond leading order, there is an overlap between central and
fragmentation region contributions, both for real and virtual
corrections. Moreover both contributions are divergent and require a
regulator. In \cite{Hentschinski:2011tz,Chachamis:2012cc} it has been
shown through the explicit calculation of NLO corrections for quark
and gluon forward jet vertices, that this overlap can be removed
through a subtraction procedure which removes from the NLO impact
factors the corresponding matrix element which contains an internal
reggeized gluon line, {\it i.e.} through subtraction of the factorized
contribution. The remaining dependence on the regulator then cancels
at the level of the NLO cross section, which combines NLO corrections
from both fragmentation and central region, see
\cite{Hentschinski:2011tz,Chachamis:2012cc, Chachamis:2012mw,
  Hentschinski:2020rfx} for a detailed discussion. In the following we
will slightly formalize this observation by introducing a transition
function, generalizing a similar object used in
\cite{Chachamis:2013hma} for the calculation of the 2-loop gluon Regge
trajectory. Defining the bare one-loop 2-reggeized-gluon Green's
function $G_B(\bk_1,\bk_2)$ through
\begin{align}
  \label{eq:Uuu}
  G_B(\bk_1, \bk_2; \rho)
& =  \delta^{(2 + 2 \epsilon)} (\bk_1 + \bk_2) +  G_B^{(1)}(\bk_1, \bk_2; \rho) + \ldots \notag \\
G_B^{(1)}(\bk_1, \bk_2; \rho)  & = \rho
V(\bk_1, \bk_2) + \frac{ \Sigma\left(\rho; \epsilon, \frac{\bk_1}{\mu^2} \right) + \Sigma^*\left(\rho; \epsilon, \frac{\bk_1}{\mu^2} \right) }{-2i \bk_1^2}\delta^{(2 + 2 \epsilon)} (\bk_1 + \bk_2),
\end{align}
we define at first the following subtracted bare NLO coefficient,
\begin{align}
  \label{eq:coeff}
   C_{a,B}^{(1)}(\bk, \rho)  & =   h_{a}^{(0)}(\bk) +  h_{a}^{(1)}(\bk, \rho)
-
 \left[h_{a}^{(0)} \otimes G^{(1)}_B(\rho) \right](\bk)
\end{align}
where we implied the following  expansion in $\alpha_s$ of the impact factors,
\begin{align}
  \label{eq:impactExpand}
   h_{a}(\bk, \rho) & =  h_{a}^{(0)}(\bk) +  h_{a}^{(1)}(\bk, \rho) + \ldots .
\end{align}
Rapidity divergences in the one-loop correction  to the impact
factors are 
understood to be regulated through lower cut-offs on the rapidity of
all particles, $\eta_i > -\rho/2$ with $\rho \to \infty$ and
$i = 1, \ldots, n$ for $n$ the number of particles produced in the
fragmentation region of the initial parton $a$. For virtual
corrections, the regularization is again implemented through tilting
light-cone directions of the high energy effective action. Finally
note that for the fragmentation of the parton $b$, the regulator would
be $\eta_i < \rho/2$ with $\rho \to \infty$. We further introduced for
this paragraph the following convolution convention
\begin{align}
  \label{eq:64}
\left[  f \otimes g\right](\bk_1, \bk_2) & \equiv  \int d^{2 + 2 \epsilon}  \bq 
f(\bk_1, \bq) g(\bq, \bk_2).
\end{align}
Ignoring terms beyond NLO accuracy and combining NLO corrections in
the fragmentation region of both  partons as well as at the central
rapidities, the partonic cross section can be compactly written
as\footnote{Note that the impact factors themselves might depend on
  additional transverse momenta; this is however irrelevant for the
  following discussion of high energy factorization and we therefore
  suppress this dependence in the following.}
\begin{align}
  \label{eq:dsigNLO}
    d \sigma_{ab}^{\text{NLO}} & = \left[ C_{a,B}(\rho) \otimes G_B(\rho) \otimes C_{b,B}(\rho) \right] .
\end{align}
As a next step we define  a renormalized Green's function $G_R$  through
\begin{align}
  \label{eq:58}
    G_B(\bk_1, \bk_2; \rho) & =  \left[  Z^+\left(\frac{\rho}{2} -\eta_a\right) \otimes G_R\left( \eta_a, \eta_b\right)  \otimes Z^- \left(\frac{\rho}{2}+ \eta_b\right) \right] (\bk_1, \bk_2) ,
\end{align}
where the transition functions $Z^\pm$ possess the following perturbative expansion
\begin{align}
  \label{eq:61}
    Z^\pm(\hat{\rho}; \bk, \bq)  & = 
             \delta^{(2 + 2 \epsilon)}(\bk - \bq) + \hat{\rho} K_{\text{BFKL}}(\bk, \bq) + f^\pm(\bk, \bq) + \ldots ,
\end{align}
and are to all orders defined through the  following BFKL equation,
\begin{align}
  \label{eq:BFKL_forZ}
  \frac{d}{d\hat{\rho}} Z^+(\hat{\rho}; \bk, \bq)  & =  
\left[  Z^+(\hat{\rho}) \otimes K_{\text{BFKL}} \right] (\bk, \bq), \notag \\
\frac{d}{d\hat{\rho}} Z^-(\hat{\rho}; \bk, \bq)  & =  
\left[ K_{\text{BFKL}} \otimes  Z^-(\hat{\rho})\right] (\bk, \bq),
\end{align}
where
\begin{align}
  \label{eq:62}
  K_{\text{BFKL}}(\bk, \bq) & =  K^{(1)}(\bk, \bq) +  K^{(2)}(\bk, \bq) + \ldots
\end{align}
denotes the still undetermined BFKL kernel;  $ f^\pm(\bk, \bq)$ 
parametrizes finite contributions and is in principle
arbitrary. Symmetry of scattering amplitudes suggests
$ f^+(\bk, \bq) = f^-(\bk, \bq)$,  while Regge theory suggests to fix it in
such a way that terms which are not enhanced by the parameter $\eta$ are entirely transferred from the renormalized
Green's function to the impact factors. Note that the factorization
parameter  $\eta$ plays a r\^ole analogous to the factorization scale
in {\it i.e.}\ collinear factorization and parametrizes the scale
ambiguity associated with high energy factorization.  Fixing the lowest order terms of $G_R$ through
\begin{align}
  \label{eq:GR_expand}
  G_R(\eta_a, \eta_b; \bk_1, \bk_2) & = \delta^{(2 + 2 \epsilon)}(\bk_1 + \bk_2)
+
 G_R^{(1)}(\eta_a, \eta_b; \bk_1, \bk_2) + \ldots,
\end{align}
and expanding the right-hand side up to linear terms, we obtain
\begin{align}
  \label{eq:fixes_us}
   K^{(1)}(\bk_1, \bk_2) & = 
V(-\bk_1, \bk_2) + \delta^{(2 + 2 \epsilon)}(\bk_1 - \bk_2)\omega^{(1)}(\bk_1),
\notag \\
 \omega^{(1)}\left(\epsilon, \frac{{\bm k}^2}{\mu^2} \right) &  = - \frac{\alpha_s \CA}{\pi \epsilon} \left(\frac{\bk}{\mu^2} \right)^\epsilon + \mathcal{O}(\epsilon),
\notag \\
f^{\pm,(1)}\left( \bk_1, \bk_2 \right) & =  \delta^{(2 + 2 \epsilon)}(\bk_1 - \bk_2) \frac{\alpha_s}{4 \pi} \left[ - \frac{1}{\epsilon} \left( \frac{5 \CA}{3} - \frac{2 n_f}{3}\right) + \frac{31 \CA}{9} - \frac{10 n_f}{9} \right] + \mathcal{O}(\epsilon).
\end{align}
As a consequence
\begin{align}
  \label{eq:GR1loop}
  G_R^{ (1)}(\eta_a, \eta_b; \bk_1, \bk_2)  & = (\eta_a - \eta_b) K^{(1)}(\bk_1, \bk_2) .
\end{align}
Using Eq.~\eqref{eq:BFKL_forZ}, it is then straightforward to show that
\begin{align}
  \label{eq:BFKL_forGR}
  \frac{d}{d\eta_a} G_{R}(\eta_a, \eta_b; \bk_1, \bk_2) & = \left[ K^{(1)}(\bk_1, \bk_2) \otimes  G_{R}(\eta_a, \eta_b)\right](\bk_1, \bk_2) \notag \\
 \frac{d}{d\eta_b} G_{R}(\eta_a, \eta_b; \bk_1, \bk_2) & = \left[ G_{R}(\eta_a, \eta_b) \otimes  K^{(1)}(\bk_1, \bk_2) \right](\bk_1, \bk_2).
\end{align}
Note that through imposing,
\begin{align}
  \label{eq:GR0}
   G_{R}(\eta_b, \eta_b; \bk_1, \bk_2) & = \delta^{(2 + 2 \epsilon)} (\bk_1 + \bk_2), & \text{and} && \eta_a& > \eta_b,
\end{align}
no over-counting occurs. With all factors fixed, we insert now Eq.~\eqref{eq:58} into the NLO cross section Eq.~\eqref{eq:dsigNLO}, which then immediately leads to
\begin{align}
  \label{eq:dsig_renom}
  d\sigma_{ab}^{\text{NLO}} = \left[C_{a,R}(\eta_a) \otimes G_R(\eta_a, \eta_b) \otimes C_{b,R}(\eta_b) \right] ,
\end{align}
where
\begin{align}
  \label{eq:CRs}
  C_{a,R}(\eta_a;\bk_1) & \equiv \left[ C_a(\rho)  \otimes Z^+\left(\frac{\rho}{2} - \eta_a\right)\right](\bk_1), \notag \\
 C_{b,R}(\eta_b;\bk_2) & \equiv \left[  Z^+\left(\frac{\rho}{2} +\eta_b\right) \otimes C_b (\rho) \right](\bk_2)  \, .
\end{align}
In the following paragraph we will provide an explicit verification of
this procedure, through applying it to the forward Higgs impact factor. For simplicity we note that the finite coefficient is at NLO given by the following general expression,
\begin{align}
  \label{eq:CNLO}
  C_{R}^{\text{NLO}}(\bk) & = h_a^{(0)}(\bk) + h_a^{(1)}(\bk) + h_a^{(0)}\otimes \left[(-\frac{\rho}{2}- \eta_a) K^{(1)} - f^{-, (1)} \right](\bk) \notag \\
&=
h_a^{(0)}(\bk) + h_a^{(1)}(\bk, \rho) -  \frac{\alpha_s N_c (\frac{\rho}{2} + \eta_a)}{\pi } \left[ \int \frac{d^{2 + 2 \epsilon}\br}{ \br^2}  h_a^{(0)}\left((\br + \bk)^2\right) \right] \notag \\
&
+h_a^{(0)}(\bk)\frac{\alpha_s}{2 \pi}\left(\frac{\bk^2}{\mu^2} \right)^\epsilon \left[\frac{ \CA(\rho +2 \eta_a)}{\epsilon}   + \frac{1}{\epsilon} \left( \frac{5 \CA}{6} - \frac{2 n_f}{6}\right) - \frac{31 \CA}{18} + \frac{10 n_f}{18} \right]
.
\end{align}

\section{The impact factor for forward Higgs production}
\label{sec:partonic}

\begin{figure}[t]
  \centering
  \parbox{5cm}{\includegraphics[width=4cm]{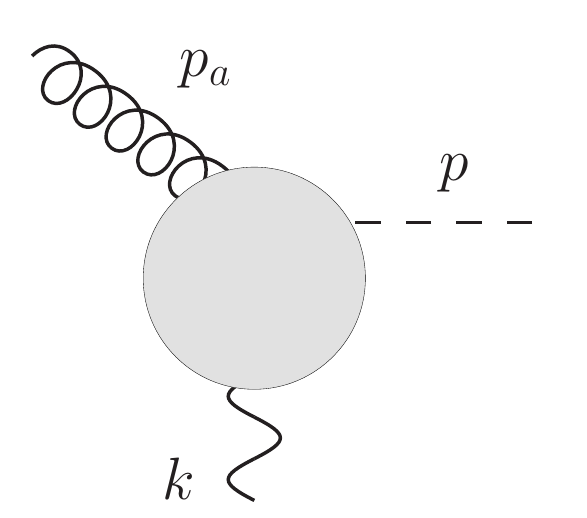}} $\qquad$ \parbox{5cm}{\includegraphics[width=4cm]{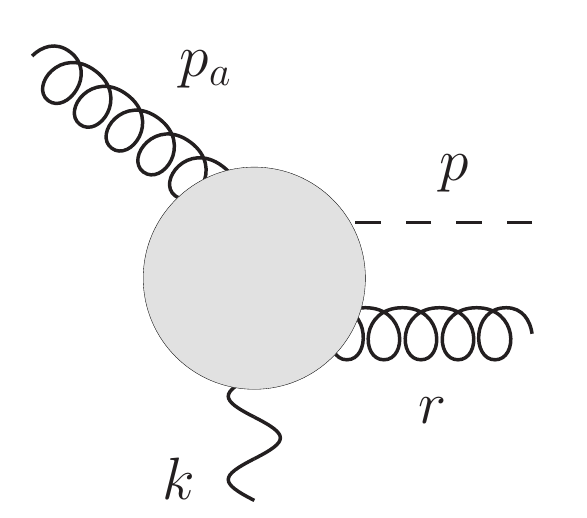}}
  \caption{Kinematics for the leading order matrix element as well as virtual corrections (left) and real next-to-leading order corrections (right). The latter implies the case where the gluon is replaced by a quark. Momenta $p_a$ and $k$ are in-going, while $p$ and $r$ are out-going.}
  \label{fig:momenta}
\end{figure}

We consider collisions of two hadrons $A$ and $B$ with momenta
$p_{A,B} = p_{A,B}^{\pm} n^\mp/2$ and squared center of mass energy $s
= p_A^+ p_B^-$ with inclusive production of an on-shell Higgs boson in
the fragmentation region of hadron $A$. The   four momentum  of the
Higgs boson $p$ and its rapidity $\eta_H$ are parametrized as
\begin{align}
  \label{eq:Higgs_momentum}
  p & =
x_H p_A  + \frac{M_H^2 + \bp^2}{x_H s}\,p_B  + p_T, 
& \eta_H & = \ln \frac{x_H p_A^+}{\sqrt{M_H^2+ \bp^2}},
\end{align}
where $p_T$ is the embedding of the Euclidean Higgs transverse momentum $\bp$ into Minkowski space, see also Fig.~\ref{fig:momenta} .

To describe the coupling of the Higgs boson to the gluonic field, we
make use of the heavy top limit and employ the following effective
Lagrangian \cite{Ellis:1975ap,Shifman:1979eb},
\begin{align}
  \label{eq:4}
  \mathcal{L}_{\text{eff}} & = - \frac{1}{4} g_H H F_{\mu\nu}^a F_a^{\mu\nu}
\end{align}
with $H$ the scalar (Higgs) field and $g_H$ the effective
coupling \cite{Dawson:1990zj, Ravindran:2002dc}
\begin{align}
  \label{eq:gH_Id}
  g_H & = -\frac{\alpha_s}{3 \pi v} \left(1 + \frac{\alpha_s}{4 \pi}
        11  \right) + \mathcal{O}(\alpha_s^3)\, . 
\end{align}
Since the top quark has been integrated out, the strong coupling
$\alpha_s$ is  evaluated for $n_f = 5$ flavors and  $v^2 =
1/(\sqrt{2}G_F)$ with $G_F$ the Fermi constant. Working under the assumption that multi-reggeized gluon exchanges can be neglected, the hadronic differential cross section is factorized into
\begin{align}
  \label{eq:1}
  \frac{d^3 \sigma}{d^2 \bp d x_H} & = \int_{x_H}^1 \frac{dz}{z} \sum_{a = {q,g}} f_a\left(\frac{x_H}{z}, \mu_F^2 \right)  \int \frac{d^2 \bk}{\pi} \frac{d \hatC_{ag^*\to H}(\mu_F^2, \eta_a; z, \bk)}{d^2 \bp d x_H} \mathcal{G}(\eta_a, \bk),
\end{align}
where $ \mathcal{G}(\eta_a, \bk)$ denotes the unintegrated gluon distribution of hadron $B$ which parametrizes  non-perturbative input of hadron $B$ and is subject to BFKL evolution; $\eta_a$ is a factorization parameter associated with the highest gluon rapidity absorbed into the unintegrated gluon density. In terms of the elements defined in the previous section we have
\begin{align}
  \label{eq:2}
   \mathcal{G}(\eta_a, \bk, Q_0) & = \int d^2 \bq\, G_R(\eta_a, \bk, \bq) h^{ugd}(\bq, Q_0),
\end{align}
where $h^{ugd}$ is obtained as the convolution of partonic impact
factor and parton distribution functions. In particular,  collinear
singularities, which arise from the infra-red region of transverse
momentum integration  are assumed to be absorbed into the parton
distribution function of hadron $B$ following the general procedure
outlined in \cite{Catani:1994sq}, see also \cite{Ciafaloni:1998gs,Ciafaloni:1998hu}. The dependence on the scale $Q_0$ is understood to arise as a consequence of such a factorization of collinearly enhanced contributions.  For the partonic differential coefficient, we assume the following perturbative expansion
\begin{align}
  \label{eq:expansion}
   \frac{d^3 \hatC^{NLO}_{ag^* \to H}}{d x_H d^2 \bp}  &= \sigma_0 \left( \frac{d^3 \hatC^{(0)}_{ag^* \to H}}{d x_H d^2 \bp} + \frac{ \alpha_s}{2 \pi}  \cdot  \frac{d^3 \hatC^{(1)}_{pg^* \to H}}{d x_H d^2 \bp} + \ldots \right). &
a&=q,g,
\end{align}
With
\begin{align}
  \label{eq:sigma0}
 \frac{d^3 {h}_{ag^* \to H}^{(0)}}{dx_H d^2 \bp} & = 
\sigma_0 \int \frac{d k^-}{k^-}   \delta^{(2)}(\bp - \bk) \delta(1-z)  \delta\left(1- \frac{M_H^2 + {\bm k}^2}{p_a^+ k^-} \right), &
       \sigma_0 & = \frac{g_H^2\pi}{8(N_c^2 -1)},
\end{align}
we have
\begin{align}
  \label{eq:coeffLO}
  \frac{d \hatC_{gg^*\to H}^{(0)}(\mu_F^2, \eta_a; z, \bk)}{d^2 \bp d x_H} 
 & =   \delta^{(2)}(\bp - \bk) \delta(1-z), 
\end{align}
at leading order, while the corresponding contribution from the
quark-channel vanishes.  In the following we will determine the
next-to-leading order corrections to this impact factor. This will be
the main result of this paper.

\subsection{Virtual next-to-leading order corrections}
\label{sec:virtual-corrections}

Virtual NLO corrections to the operator $-\frac{1}{2} \tr [G_{\mu\nu}G^{\mu\nu}]$ have been calculated in \cite{Nefedov:2019mrg}. Adapting the conventions of that paper to the ones used here we find:
\begin{align}
  \label{eq:hKT_virtual}
  \frac{d {h}_{gg^* \to H}^{(1)}(z, \bk)}{d^2 \bp d x_H}  & =  \frac{d {h}^{(0)}_{gg^*\to H}(z, \bk)}{d^2 \bp d x_H}  
\frac{\alpha_s}{2 \pi}  \cdot  \left(\frac{\bk^2}{\mu^2} \right)^\epsilon  \bigg\{ -\frac{\CA}{\epsilon^2}  - \frac{1}{\epsilon} \left(\frac{8 \CA}{3} - \frac{2 n_f}{3} \right) \notag \\
& \hspace{-2cm}+ \frac{\CA}{\epsilon}  \left[-\rho + \ln\frac{{\bm k}^2}{(p_a^+)^2}   \right]
 + \CA \left[ 2 \text{Li}_2 \left(1 + \frac{M_H^2}{\bk^2} \right)  +  \frac{\pi^2 }{6} + \frac{49}{9}\right]  + 11  - \frac{10}{9}n_f
 \bigg\}\notag \\
\end{align}
where we also added the contribution due to the 1-loop corrections  of the Higgs-gluon-gluon coupling in the heavy quark limit, Eq.~\eqref{eq:gH_Id}. 

\begin{figure}[t]
 \parbox{2.2cm}{
\includegraphics[width=2.2cm]{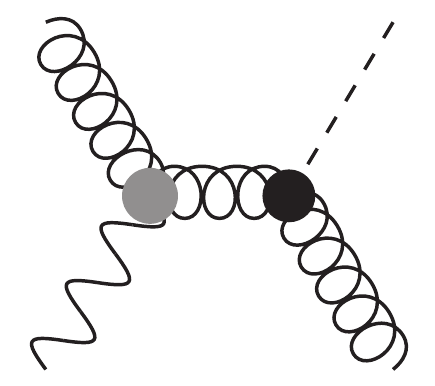}
} 
 \parbox{2.2cm}{
 \includegraphics[width=2.2cm]{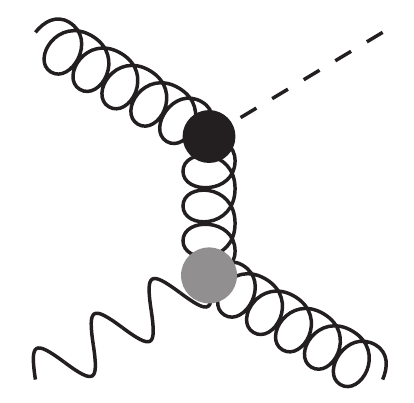}
} 
\parbox{2.2cm}{
\includegraphics[width=2.2cm]{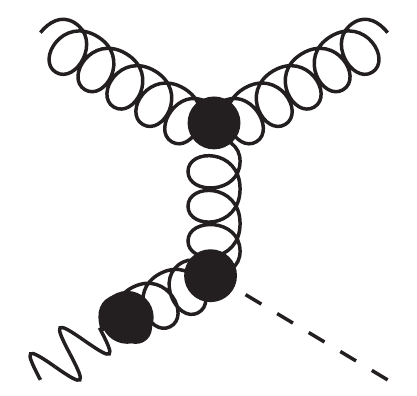}
} 
\parbox{2.2cm}{
\includegraphics[width=2.2cm]{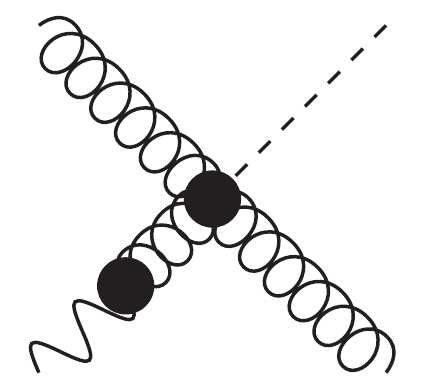} 
} \hspace{2cm}
\parbox{2.2cm}{
\includegraphics[width=2.2cm]{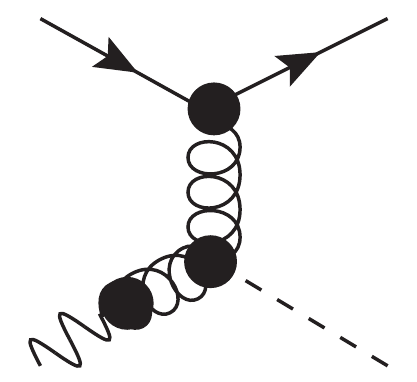}
} 
\\
\vspace{.5cm}

 \parbox{2cm}{\includegraphics[width=2cm]{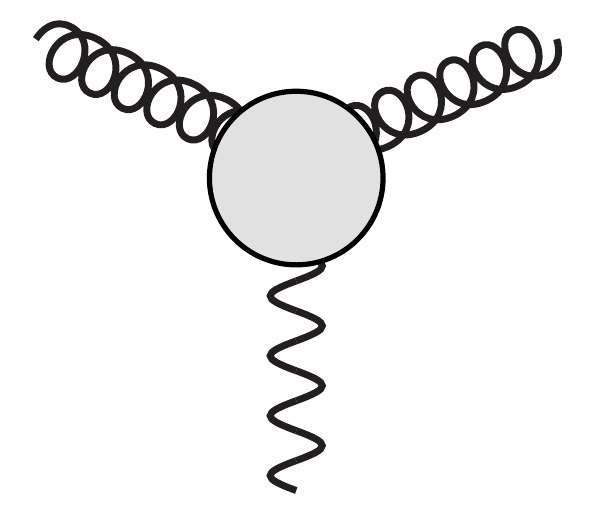}}
  =
  \parbox{2cm}{\includegraphics[width=2cm]{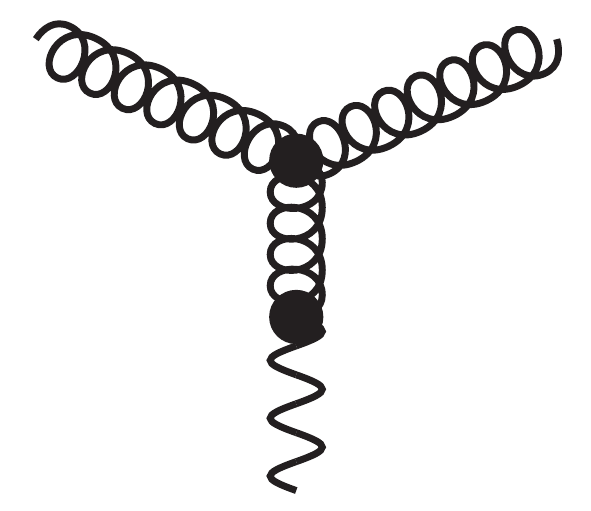}} 
  +
  \parbox{2cm}{\includegraphics[width=2cm]{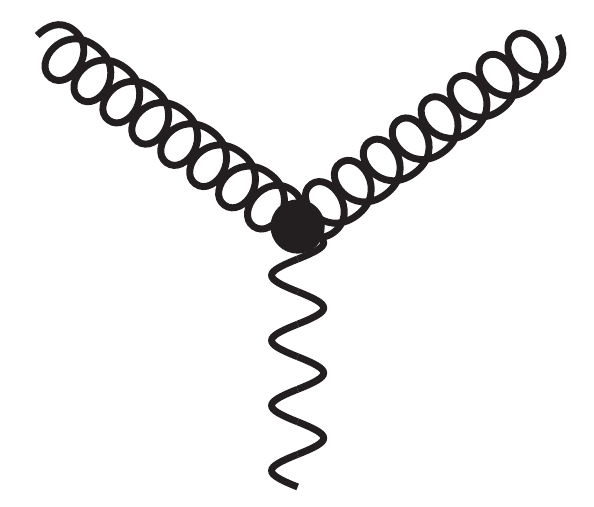}} 
  \caption{\it  Feynman diagrams for real correction in  the gluon (left) and quark (right) channels respectively.  Here the wavy line indicates the off-shell reggeized gluon state. The gray vertex indicates the gluon-gluon- reggeized gluon which is obtained as a combination of a first order induced vertex and the three-gluon vertex, see second line. The quark channel includes both contributions due to quarks and anti-quarks. }
  \label{fig:diagrams}
\end{figure}

\subsection{Real next-to-leading order corrections}
\label{sec:real-corrections}

Real NLO corrections contain both contributions from the gluon and the quark channel. The relevant Feynman diagrams are depicted in Fig.~\ref{fig:diagrams}. Our convention for momenta is as follows
\begin{align}
  \label{eq:5}
  g(p_a) + r_+(k)&  \to H(p) + g(r),
\end{align}
where we replace $g \leftrightarrow q, \bar{q}$ for the contributions with initial and final (anti-) quark states. We further use  $0< z<1$ to parametrize the initial parton momentum fraction, carried on by the Higgs particle.  For the gluon  channel we find
\begin{align}
  \label{eq:hadronic_G}
  \frac{d^3{h}_{gg^* \to Hg}^{(0)}(z, \bk)}{dx_H d^2 \bp}  & = \frac{\alpha_s \CA \sigma_0}{2 \pi_\epsilon  \bk^2} \, H_{ggH}(z, \bp, \bk) \theta\left(\eta_g + \frac{\rho}{2}\right)
\end{align}
where $\eta_g = \ln \frac{(1-z)x_H p_A^+}{z \sqrt{\br^2}}$ is the gluon rapidity and $\rho$ the regulator for the high energy divergence, which we take  in the limit $\rho \to \infty$. We further  kept the dependence the dimensional regularization parameter $\epsilon$ explicit:
\begin{align}
  \label{eq:Hdef}
  H_{ggH}&(z, \bp, \bk) 
 =
    \frac{2}{z(1-z)} \bigg\{2z^ 2
+ \frac{(1-z)z M_H^2 ({\bm k}\cdot {\bm r})[z^2 + (1-z) \cdot 2 \epsilon] -
 2 z^3 ({\bm p}\cdot {\bm r }) ({\bm p}\cdot {\bm k })}{{\bm r}^2 ({\bm p}^2 + (1-z)M_H^2)}
\notag \\
&+ \frac{(1 + \epsilon) (1-z)^2 z^2 M_H^4}{2} \left(\frac{1}{{\bm \Delta}^2 + (1-z) M_H^2} +  \frac{1}{{\bm p}^2 + (1-z) M_H^2} \right)^2 
\notag \\
& 
- \frac{2 z^2 ({\bm p} \cdot {\bm \Delta})^2 + 2 \epsilon \cdot (1-z)^2 z^2 M_H^4 }{({\bm p}^2 + (1-z) M_H^2)({\bm \Delta}^2 + (1-z) M_H^2)} 
- \frac{2 z(1-z)^2 M_H^2}{{\bm \Delta}^2 + (1-z) M_H^2} -  \frac{2 z(1-z)^2 M_H^2}{{\bm p}^2 + (1-z) M_H^2}
\notag \\
& 
- \frac{(1-z)z M_H^2 ({\bm k}\cdot {\bm r})[z^2 + (1-z) \cdot 2 \epsilon] -
 2 z^3 ({\bm \Delta}\cdot {\bm r }) ({\bm \Delta}\cdot {\bm k })}{{\bm r}^2 ({\bm \Delta}^2 + (1-z) M_H^2)} \bigg\}
\notag \\
&  + 
\frac{2\bk^2}{\br^2}\left\{\frac{z}{1-z} + z(1-z)
 +
 2 (1 + \epsilon) \frac{(1-z)}{z} \frac{({\bm k}\cdot {\bm r})^2}{{\bm k}^2 {\bm r}^2}\right\};
\end{align}
 $\br = \bk - \bp$ denotes the transverse momentum of the real final state parton and 
\begin{align}
  \label{eq:6}
 {\bm \Delta} = z \br  -(1-z) \bp . 
\end{align}
For the case of an initial quark we obtain instead
 \begin{align}
  \label{eq:hadronic_Q}
  \frac{d^3 {h}_{qg^* \to Hq}^{(0)}(z, \bk)}{dx_H d^2 \bp}  & = \frac{\alpha_s \CF \sigma_0}{ 2 \pi_\epsilon  \bk^2} \, H_{qqH}(z, \bp, \bk),
\end{align}
with
\begin{align}
  \label{eq:7}
   H_{qqH}(z, \bp, \bk) & = \frac{1+\epsilon}{z} \left[z^2 +4 (1-z)  \frac{(\bk \cdot \br)^2}{\bk^2 \br^2}\right].
\end{align}
Both Eq. (\ref{eq:Hdef}) and Eq. (\ref{eq:7}) were evaluated directly
using Lipatov's high energy effective action as well as using the
conventional $k_T$ factorization procedure, where the
sum over polarization of the incoming off-shell gluon is given by eikonal projectors. We furthermore  cross-checked the result  numerically using \KaTie~\cite{vanHameren:2016kkz}. \\

Note that the quark channel is free of high energy divergences and we
therefore took already the limit $\rho \to \infty$. To address the
high energy divergence at $z=1$ of the gluonic real corrections, we
note at first that
\begin{align}
  \label{eq:8}
  \lim_{z\to 1} \frac{\alpha_s \CA \sigma_0}{2 \pi_\epsilon}\frac{H_{ggH}(z, \bp, \bk)}{\bk^2} & = \frac{\alpha_s \CA \sigma_0}{ \pi_\epsilon \cdot \br^2} \frac{1}{1-z} + \text{finite}\, ,
\end{align}
where `finite' indicates all the terms which do not require a regulator. Making use of the following identity, where   $f(z)$  is a generic test function,
\begin{align}
  \label{eq:ID}
  \lim_{\rho \to \infty} \int_{x_H}^1  dz\frac{f(z) \theta(\eta_g + \rho/2)}{1-z} & = 
\int_{x_H}^1  dz\frac{f(z)}{(1-z)_+} + f(1)\left[\ln\frac{x_Hp_A^+}{\sqrt{\br^2}} + \frac{\rho}{2}  \right],
\end{align}
with
\begin{align}
  \label{eq:plusdef}
  \int_{x_H}^1 dz \frac{f(z)}{(1-z)_+} & =  \int_{x_H}^1 dz \frac{f(z) - f(1)}{1-z} - \int_0^{x_H} dz \frac{f(1)}{1-z} ,
\end{align}
we identify the high energy singularity of the real corrections as
\begin{align}
  \label{eq:HggH_rapdiv}
 \frac{\alpha_s \CA \sigma_0}{2 \pi_\epsilon}\frac{ H_{ggH}(z, \bp, \bk)}{\bk^2} & = \delta(1-z)\frac{\rho}{2} \sigma_0 \cdot \frac{\alpha_s \CA}{\pi_\epsilon \br^2} + \ldots,
\end{align}
while the $z\to 1$ singularity in $H_{ggH}$ is now regulated through a plus-prescription and the dots indicate terms finite in the limit $\rho \to \infty$. 

\subsection{Counter-terms}
\label{sec:counterterms}

Our result requires a number of counter-terms both to ultra-violet renormalization, collinear factorization (initial parton) and high energy factorization (reggeized gluon field). For the former two we will employ the $\overline{\text{MS}}$-scheme, wile for the latter we will make use of the scheme presented in Sec.~\ref{sec:subtrans}. The ultra-violet counter-term is identical to the one used in the determination of collinear NLO corrections in \cite{Dawson:1990zj} and can be entirely expressed through renormalization of the QCD strong coupling in the Higgs-gluon-gluon coupling constant $g_H$. With
\begin{align}
  \label{eq:alphaS_renom}
  \alpha_s = \alpha_s(\mu_R)\left[1 + \frac{\alpha_s(\mu_R) \beta_0}{(4 \pi)} \left(\frac{1}{\epsilon} + \ln \frac{\mu_R^2}{\mu^2} \right) \right], &\,\,\,\,\,\,
\beta_0 = \frac{11\CA}{3} - \frac{2n_f}{3},
\end{align}
where in  the following we set $\mu_R = \mu$.  
The remaining singularities both in the limit $\epsilon \to 0$ and $\rho \to \infty$ can be factored into process independent functions associated with the external legs (collinear parton and off-shell reggeized gluon state). With
\begin{align}
  \label{eq:crossi}
   \frac{d^3 {h}_{ag^* \to Ha}}{dx_H d^2 \bp} & =   \frac{d^3 {h}_{ag^* \to H}^{(0)}}{dx_H d^2 \bp}  +   \frac{d^3 {h}_{ag^* \to H}^{(1)}}{dx_H d^2 \bp}  +  \frac{d^3 {h}_{ag^* \to Ha}^{(0)}}{dx_H d^2 \bp} ,
\end{align}
the physical coefficient is then implicitly defined through the relation
\begin{align}
  \label{eq:coeff_renom}
  \frac{d^3 {h}^{(0)}_{ag^* \to Ha}}{dx_H d^2 \bp} & =  \sum_{b=q,g} \int_{x_H}^1 d\xi \int d^{2 + 2 \epsilon} \tilde{\bk} \,  \frac{d^3 \hat{C}_{bg^* \to Hb}(z, \tilde{\bk}, \bp)}{dx_H d^2 \bp}  \Gamma_{ba}\left(\xi, \frac{\mu_F^2}{\mu^2}\right) \tilde{\Gamma}_{g^*g^*}(\xi, \tilde{\bk}, \bk),&
\end{align}
where
\begin{align}
  \label{eq:1loopg}
  \Gamma_{ba}(z) & = \delta_{ba}\delta(1-z) -
\frac{\alpha_s}{2 \pi} \left(\frac{1}{\epsilon} + \ln \frac{\mu_F^2}{\mu^2} \right)P_{ga}(z) + \mathcal{O}(\alpha_s^2), &
a& = q, g
\end{align}
are   the 1-loop partonic parton distribution   with  gluon splitting functions,
\begin{align}
  \label{eq:splitting}
  P_{gq}(z) & = \CF\frac{1 + (1-z)^2}{z},
\notag \\
P_{gg}(z)  &= 2 \CA \left[\frac{z}{(1-z)_+} + \frac{1-z}{z} + z(1-z) \right] + \frac{\beta_0}{2} \delta(1-z),
\end{align}
where corresponding quark splitting functions are absent since the leading order quark coefficient vanishes. 
The 1-loop unintegrated gluon distribution is, following Sec.~\ref{sec:subtrans}, given by
\begin{align}
  \label{eq:Gammatilde}
   \tilde{\Gamma}_{g^*g^*}(\xi, \tilde{\bk}, \bk) & = \delta\left(1-\xi \right)  \delta^{(2)}(\bk - \tilde{\bk})  \bigg[
1-
\frac{\alpha_s}{2 \pi}\left(\frac{\bk^2}{\mu^2} \right)^\epsilon \left( \frac{5 \CA - 2n_f}{6 \epsilon}  - \frac{31 \CA - 10 n_f}{18}  \right)
\notag \\
& \qquad +  \delta\left(1-\xi \right) (\rho + 2 \eta_a)
 \left[ \frac{\alpha_s \CA}{2 \pi_\epsilon (\tilde{\bk} - \bk)^2} -
 \delta^{(2)}(\bk - \tilde{\bk})\frac{\alpha_s}{2 \pi \epsilon} \left(\frac{\bk^2}{\mu^2} \right)^\epsilon  \right],
\end{align}
which contains apart from the 1-loop BFKL kernel in the second line, also terms due to the gluon self-energy. In the current  setup we preferred to formulate this distribution in $d = 4 + 2\epsilon$, while it is in principle straightforward to remove the remaining UV divergence through an appropriate counter term associated with the gluon self-energy.  

\subsection{Subtraction mechanism to achieve numerical stability}
\label{sec:subtraction_num}
Given the above counter terms, it is a relatively straightforward task
to verify finiteness of the resulting coefficient in the limit
$\rho \to \infty$ and $\epsilon \to 0$. While the subtraction of high
energy and ultraviolet singularities is straightforward, extracting of
infrared singularities is more cumbersome and requires the use of
phase space slicing parameters, see
\cite{Hentschinski:2014lma,Hentschinski:2014bra,Hentschinski:2014esa}
for instance. While this is sufficient to demonstrate finiteness at a
formal level, the use of such phase space slicing parameters is in
general complicated for numerical studies at NLO accuracy. For the
case of collinear NLO calculation, the by now conventional tool to
overcome this difficulty is provided by subtraction methods, in
particular the dipole subtraction as formulated in
\cite{Catani:1996vz}. Within the current setup, collinear and soft
singularities are directly associated with the convolution integral
over transverse momenta and the formulas of \cite{Catani:1996vz}
cannot be directly translated to the present case. In the following we
therefore present a subtraction mechanism which closely follows the
spirit of \cite{Catani:1996vz}, but which is adapted to the current setup. In
particular a generalization to other partonic high energy coefficients
appears to be possible. Following \cite{Catani:1996vz}, the basic idea
is to subtract a certain auxiliary term from the real NLO corrections
which a) renders the latter finite and b) can be easily integrated
analytically and added to the virtual NLO corrections.  We therefore propose the following decomposition:
\begin{align}
  \label{eq:subtR2}
   \int \frac{d^{2+2\varepsilon} \br}{\pi^{1+\varepsilon}} 
   \,\frac{\kappa(\br)}{\br^2}\, 
   G((\bp + \br)^2) 
& = 
  \int\frac{d^{2} \br}{\pi }
   \PlusDist{\kappa(\br)}G((\bp + \br)^2) \;+\; \mathcal{O}(\varepsilon)
\notag\\&\hspace{15.2ex}
  +\int \frac{d^{2+2\varepsilon}\br}{\pi^{1+\varepsilon}}
   \,\frac{\kappa(\br)}{\br^2}\,
   \frac{\bp^2G(\bp^2)}{\br^2 + (\bp + \br)^2}
\;\;, 
\end{align}
with
\begin{align}
  \label{eq:plusrp}
   \int \frac{d^{2} \br}{\pi} 
   \PlusDist{\kappa(\br)} 
   G((\bp + \br)^2) 
& \equiv 
\int\frac{d^{2}\br}{\pi}
 \,\frac{\kappa(\br)}{\br^2}\, \left[G((\bp + \br)^2) - \frac{\bp^2 G(\bp^2)}{\br^2 + (\bp + \br)^2} \right]
\;\;.
\end{align}
The expression in the squared brackets on the right-hand side vanish in the limit $|{\bm r}| \to 0$, and $G(\bk)$ is a function which parametrizes the transverse momentum dependence of the reggeized gluon state.
The function $\kappa(\br)$ is such that the integral on the right hand side of \Equation{eq:plusrp} is well-defined, which in practice means that it does not behave worse than $\ln|\br|$ for $|\br|\to0$ and $|\br|\to\infty$.
 Furthermore, it should be such that the integral in the second line of \Equation{eq:subtR2} can be calculated analytically. Note that the factor ${\bm p}^2/ [{\bm r}^2 + ({\bm p} + {\bm r})^2]$ is needed to achieve convergence in the ultraviolet. We have the following results for the choices $\kappa(\br)=1$, for $\kappa(\br)=-\ln\br^2$, and for $\kappa(\br)=(\bp\!\cdot\!\br)^2/\br^2$ which we will need:
\begin{align}
  \label{eq:int1}
  \int \frac{d^{2 + 2 \epsilon}\br}{\pi^{1 + \epsilon}}
  \frac{1}{\br^2 [\br^2 + (\bp + \br)^2]} & = \frac{\Gamma(1-\epsilon)\Gamma^2(\epsilon)}{2 \Gamma (2\epsilon)} (\bp^2)^{\epsilon -1}
  =
  \frac{\Gamma(1-\epsilon)}{\epsilon}  (\bp^2)^{\epsilon -1} + \mathcal{O}(\epsilon)
\notag \\
 \int \frac{d^{2 + 2 \epsilon}\br}{\pi^{1 + \epsilon}}
  \frac{-\ln\br^2}{\br^2 [\br^2 + (\bp + \br)^2]}
 & = 
  \frac{\Gamma(1-\epsilon)}{(\bp^2)^{1 - \epsilon}} \left(\frac{1}{\epsilon^2} - \frac{\ln \bp^2}{\epsilon} - \frac{\pi^2}{4} \right) + \mathcal{O}(\epsilon)
\notag\\
  \int \frac{d^{2 + 2 \epsilon} \br}{\pi^{1 + \epsilon}} 
  \frac{\left(\bp\cdot \br\right)^2}{(\br^2)^2[\br^2+(\bp + \br)^2]} 
  & = 
  \frac{\Gamma(1-\epsilon)}{(\bp^2)^{-\epsilon}} \left(\frac{1}{2 \epsilon} - 1 + \ln 2 \right) + \mathcal{O}(\epsilon) .
\end{align}
Note that through representing the   function $G(\bk^2)$ through its
Mellin transform, it is  further possible to verify  that the
subtraction does not generate any divergent left-overs,  see
Appendix~\ref{sec:finiteness} for an explicit calculation. 
 Following \cite{Catani:1996vz}, we further use in the subtraction
 terms the splitting functions which are not averaged over the regarding azimuthal angle. They are given by
\begin{align}
  \label{eq:Pgg}
   \hat{P}_{gg}^r(z, \bp, \br) & =  2\CA \left(\frac{z}{(1-z)_+} + (1-z)z
+ 2 (1 + \epsilon) \frac{1-z}{z} 
\frac{(\bp \cdot \br)^2}{\bp^2 \br^2} \right) \notag \\
\hat{P}_{qg}^r(z, \bp, \br) & =
\frac{\CF}{z} \left(z^2 + 4 (1-z) \frac{(\bp \cdot \br)^2}{\bp^2 \br^2}  \right),
\end{align}
and yield after averaging over the azimuthal angle the real part of the regarding splitting functions in $d = 4 + 2 \epsilon$ dimensions, {\it i.e.}
\begin{align}
  \label{eq:average}
  \int \frac{d^{1+ \epsilon} \Omega}{\pi^{1 + \epsilon}}\hat{P}^r_{ag}(z, \bp, \br) &= P^{r}_{ag}(z, \epsilon) & a& = q,g,
\end{align}
where
\begin{align}
  \label{eq:d_dim_plitings}
   P^{r}_{qg}(z, \epsilon) & = \CF\frac{1 + (1-z)^2 + \epsilon z^2}{z}
\notag \\ P^{r}_{gg}(z, \epsilon) & = 2\CA \left(\frac{z}{1-z} + \frac{1-z}{z} + z(1-z) \right)
\end{align}
We finally have
\begin{align}
  \label{eq:subtR2XX}
   \int \frac{d^{2 + 2 \epsilon} {\bm r}}{\pi^{1+\epsilon} {\bm r}^2} G\left(({\bm p} + {\bm r})^2\right) \hat{P}_{gg}^r(z, \br, \bk) 
& = \left[ P_{gg}^r(z,0) \frac{\Gamma(1-\epsilon)}{\epsilon (\bp^2)^{-\epsilon}} + \frac{1-z}{z} 2\left(\ln 2 -\frac{1}{2} \right)  \right] G({\bm p}^2) 
\notag \\
& +
   \int \frac{d^{2} {\bm r}}{\pi}
   \PlusDist{\hat{P}_{gg}^r(z, \br, \bk)}G(({\bm p} + {\bm r})^2) ,
\notag \\
 \int \frac{d^{2 + 2 \epsilon} {\bm r}}{\pi^{1+\epsilon} {\bm r}^2} G\left(({\bm p} + {\bm r})^2\right) \hat{P}_{qg}^r(z, \br, \bk) 
& = \left[ P_{qg}^r(z,0) \frac{\Gamma(1-\epsilon)}{\epsilon (\bp^2)^{-\epsilon}} + \frac{1-z}{z} 2\left(\ln 2 -1 \right)  \right] G({\bm p}^2) 
\notag \\
& +
   \int \frac{d^{2} {\bm r}}{\pi}
   \PlusDist{\hat{P}_{qg}^r(z, \br, \bk)}G(({\bm p} + {\bm r})^2) .
\end{align}
While for the term associated with the  high energy divergence  we further need
\begin{align}
  \label{eq:also}
    \int \frac{d^{2 + 2 \epsilon} {\bm r}}{\pi^{1+\epsilon} {\bm r}^2} 
 \ln\frac{x_Hp_A^+}{|\br|}  G\left(({\bm p} + {\bm r})^2\right) 
& = 
\frac{\Gamma(1-\epsilon)}{ (\bp^2)^{-\epsilon} } \left(\frac{1}{2\epsilon^2} + \frac{1}{\epsilon} \ln\frac{x_Hp_A^+}{|\bp|}  - \frac{\pi^2}{8} \right)
\notag \\
& +
   \int \frac{d^{2} {\bm r}}{\pi}
   \PlusDist{\ln(x_Hp_A^+)-\ln|\br|}G(({\bm p} + {\bm r})^2) ,
\end{align}
where we systematically neglect terms of the order $\epsilon$ and higher. 

\subsection{The NLO  coefficient for forward Higgs production}
\label{sec:result}
We present our result for the differential cross section,
\begin{align}
  \label{eq:dsig}
  \frac{d^3 \sigma}{d x_H d^2 \bp} & = \int \frac{d^2 \br}{\pi} \int_{x_H}^1 \frac{dz}{z} \sum_{a=q,g} f_a\left(\frac{x_H}{z}, \mu_F^2  \right) \frac{d^3 \hatC^{NLO}_{ag^* \to H}}{d x_H d^2 \bp} \mathcal{G}\left(\eta_a, (\br + \bp)^2 \right)
\end{align}
where  $\mathcal{G}(\eta_a, \bk)$ is the transverse momentum dependent  unintegrated gluon density with $\bk$ the transverse momentum and $\eta_a$ the evolution parameter. We have
\begin{align}
  \label{eq:formal_def}
   \frac{d^3 \hatC^{NLO}_{ag^* \to H}}{d x_H d^2 \bp}  &= \sigma_0 \left( \frac{d^3 \hatC^{(0)}_{ag^* \to H}}{d x_H d^2 \bp} + \frac{ \alpha_s}{2 \pi}  \frac{d^3 \hatC^{(1)}_{g^* \to H}}{d x_H d^2 \bp}  + \ldots\right),
&
a & = q, g
\end{align}
with
\begin{align}
  \label{eq:LO}
     \frac{d^3 \hatC^{(0)}_{gg^* \to H}}{d x_H d^2 \bp} & =    \delta^{(2)}(\br) \delta(1-z),
&
    \frac{d^3 \hatC^{(0)}_{qg^* \to H}}{d x_H d^2 \bp} & =  0,
\end{align}
and
\begin{align}
  \label{eq:coefficent}
 &  \frac{d^3 \hatC^{(1)}_{gg^* \to H}(z, \br, \bp; \eta_a, \mu_F, \mu)}{d x_H d^2 \bp} =\notag \\
& 
 \quad =\delta(1-z)\Bigg\{ 2\CA\left[\left(\ln \frac{(1-x_H)p_A^+}{\sqrt{\br^2}} - \eta_a \right)\frac{1}{\br^2}
 \right]_+  +  \delta^{(2)}(\br) \bigg[
11 -  \frac{\beta_0}{2} \ln \frac{\bp^2}{\mu^2} - \frac{5 n_f}{9}
\notag \\
& \quad +  \CA \bigg(\frac{67}{18}  + 2 \text{Li}_2 \left(1 + \frac{M_H^2}{\bp^2} \right) - \frac{\pi^2}{12}  \bigg)
  \bigg] \Bigg\}
 +
 \frac{H^{fin.}_{ggH}(z, \bp, \br)}{ \bk^2}   +  \PlusDist{\hat{P}_{gg}^r(z, \bp, \br) }       
 \notag \\
& \quad
 +   \delta^{(2)}(\br)   \bigg[\CA\frac{1-z}{z}  \left(4\ln 2 -2 \right)
 - \ln \frac{\mu_F^2}{\mu^2} P_{gg}(z)
\bigg]   
\end{align}
while
\begin{align}
  \label{eq:quark}
&  \frac{d^3 \hatC^{(1)}_{qg^* \to H}(z, \br, \bp; \eta_a, \mu_F, \mu)}{d x_H d^2 \bp} =
 \delta^{(2)}(\br)  \bigg[  \CF \frac{1-z}{z}(4\ln 2 -4)
            - \ln \frac{\mu_F^2}{\mu^2} P_{qq}(z)
\bigg]
\notag \\
& \hspace{4cm}
+ \frac{H^{fin.}_{qqH}(z, \bp, \br)}{ \bk^2}    \PlusDist{\hat{P}_{qg}^r(z, \bp, \br) }.
\end{align}
Furthermore
\begin{align}
  \label{eq:Hfin}
  H^{fin.}_{ggH}(z, \bp, \br) & =  \frac{2\CA}{z(1-z)} \bigg\{ \frac{ (1-z)^2 z^2 M_H^4}{2} \left(\frac{1}{{\bm \Delta}^2 + (1-z) M_H^2} +  \frac{1}{{\bm p}^2 + (1-z) M_H^2} \right)^2 
\notag \\
& \quad
- \frac{2 z^2 ({\bm p} \cdot {\bm \Delta})^2  }{({\bm p}^2 + (1-z) M_H^2)({\bm \Delta}^2 + (1-z) M_H^2)} 
- \frac{2 z(1-z)^2 M_H^2}{{\bm \Delta}^2 + (1-z) M_H^2}
\notag \\
& \quad
 -  \frac{2 z(1-z)^2 M_H^2}{{\bm p}^2 + (1-z) M_H^2} - \frac{(1-z)z M_H^2 ({\bm k}\cdot {\bm r})z^2  -
 2 z^3 ({\bm \Delta}\cdot {\bm r }) ({\bm \Delta}\cdot {\bm k })}{{\bm r}^2 ({\bm \Delta}^2 + (1-z) M_H^2)}
\notag \\
& \quad
+ \frac{(1-z)z M_H^2 ({\bm k}\cdot {\bm r})z^2  -
 2 z^3 ({\bm p}\cdot {\bm r }) ({\bm p}\cdot {\bm k })}{{\bm r}^2 ({\bm p}^2 + (1-z)M_H^2)}    +  2 z^2 \bigg\} 
\notag \\
& \quad  +4\CA \frac{\bk^2}{\br^2}\left[
   \frac{(1-z)}{z} \frac{({\bm k}\cdot {\bm r})^2}{{\bm k}^2 {\bm r}^2} -    \frac{(1-z)}{z} \frac{({\bm p}\cdot {\bm r})^2}{{\bm p}^2 {\bm r}^2}  \right] \notag \\
 H^{fin.}_{qqH}(z, \bp, \br) & = 4\CF \frac{\bk^2}{\br^2}\left[
   \frac{(1-z)}{z} \frac{({\bm k}\cdot {\bm r})^2}{{\bm k}^2 {\bm r}^2} -    \frac{(1-z)}{z} \frac{({\bm p}\cdot {\bm r})^2}{{\bm p}^2 {\bm r}^2}  \right]
\end{align}
and
\begin{align}
  \label{eq:momenta}
  {\bm \Delta} & = z \br -(1- z) \bp & \bk & = \bp + \br .
\end{align}

\subsection{Scale setting}
\label{sec:scale}

As pointed out already in Sec.~\ref{sec:subtrans}, our result for the NLO coefficient depends, apart from the renormalization scale $\mu$ and the collinear factorization scale $\mu_F$ on the high energy factorization parameter $\eta_a$. For the differential cross section, this $\eta_a$ dependence is  -- at NLO accuracy -- canceled against a similar dependence in the BFKL gluon Green's function. While introducing the parameters $\eta_a$ and $\eta_b$ is natural from the point of view of the high energy factorized matrix elements, in practical applications it is more convenient to parametrize $\eta_{a,b}$ in terms of the so-called reggeization scale $s_0$, originally introduced in \cite{Fadin:1998fv}, which suggests to define
\begin{align}
  \label{eq:eta_ab}
  \eta_{a} & = \ln\frac{p_A^+}{\sqrt{s_0}} &   \eta_{b} & =- \ln\frac{p_B^-}{\sqrt{s_0}}
\end{align}
such that
\begin{align}
  \label{eq:eta}
  \eta_a - \eta_b & = \ln \frac{s}{s_0} & s & = p_A^+ p_B^-.
\end{align}
While the reggeization scale is in principle arbitrary, it is naturally chosen of the order of magnitude of a typical scale of one or  both impact factors. Note that  at NLO,  an asymmetric scale choice, {\it i.e.}\ choosing $s_0$ to be of the order of a typical scale of \emph{one} of the impact factors, leads to a modification of the NLO BFKL kernel, see {\it e.g.}\ \cite{Fadin:1998py,Bartels:2006hg}. A discussion of this scale setting appears in principle possible using the formalism introduced in Sec.~\ref{sec:subtrans}, but is clearly beyond the scope of the present paper. 

\section{Conclusion}
\label{sec:concl}

We presented the NLO corrections to the impact factor for forward
production of a Higgs boson. Our result can be used both for studies
of forward Higgs production as well as studies of Higgs-jet
configuration with jet and Higgs boson separated by a large difference
in rapidity, see {\it e.g.} \cite{Celiberto:2020tmb}. \\

Apart from the actual determination of the NLO coefficient, we further provide a prescription on how to determine such a NLO coefficient from corresponding NLO matrix elements, making use of the framework provided by Lipatov's high energy effective action. This framework turn out to be of particular use in this context for various reasons. It not only provides a manifestly gauge invariant definition of off-shell matrix elements both at tree level (which implies NLO real corrections) and  1-loop (NLO virtual corrections), but also allows for a straightforward identification of factorizing contributions, which provides then the basis for the  definition of the NLO coefficient. Nevertheless it is important to stress that the same results can be always obtained through a study of exact QCD matrix elements in the high energy limit, which is actually a necessary requirement. While  in some cases this might be even  the preferred path to determine NLO corrections,  the high energy effective action provides a natural framework to organize these results into reggeized gluon Green's function and NLO coefficient and therefore to properly define the latter.  \\

Apart from the definition of the NLO coefficient, we further proposed a subtraction mechanism to
achieve a numerically stable cancellation of soft-collinear
divergences between real, virtual corrections as well as collinear
counter-terms, along the lines of \cite{Catani:1996vz}. We believe
that such a subtraction formalism will be highly beneficial for the
numerical implementation of this and other NLO results within high
energy factorization. \\

From a formal point of view, the result will be useful to study
further the resummation of soft-collinear logarithms within high
energy factorization, see {\it
  i.e.}\cite{Sun:2011iw,Mueller:2013wwa,Xiao:2018esv, Nefedov:2020ecb} as well as the
proper definition of evolution equations for transverse momentum
dependent evolution kernels along the lines
of\cite{Gituliar:2015agu,Hentschinski:2016wya, Hentschinski:2017ayz}. 

\subsection*{Acknowledgments}
MH gratefully acknowledges support by Consejo Nacional de Ciencia y Tecnolog{\'\i}a grant number A1 S-43940 (CONACYT-SEP Ciencias B{\'a}sicas). KK acknowledges the support by the Polish National Science Centre with the grant
no.\ DEC-2017/27/B/ST2/01985.
AvH is partially supported by the Polish National Science Centre grant no.\ 2019/35/B/ST2/03531.

\appendix

\section{Finiteness of the subtraction mechanism}
\label{sec:finiteness}
To verify further that the subtraction does not generate any divergent left-overs, it is possible to represent the function $G({\bm k}^2)$ through its Mellin transform\footnote{This is the correct transverse momentum dependence for an unintegrated gluon distribution and/or a BFKL Green's function}:
\begin{align}
  \label{eq:GMell}
  G({\bm k}^2) & = \frac{1}{{\bm k}^2}\int \frac{d \gamma}{2 \pi i} \left(\frac{{\bm k}^2}{Q_0^2} \right)^\gamma \tilde{G}(\gamma),
\end{align}
where $Q_0$ is a characteristic scale of the momentum distribution. One  finds:
\begin{align}
  \label{eq:MellinCheck}
         \int \frac{d^{2 + 2 \epsilon}{\bm r}}{\pi^{1 + \epsilon}}
  \frac{1}{{\bm r}^2  [({\bm p} + {\bm r})^2]^{1-\gamma}}          
&=
\frac{\Gamma(1-\epsilon)}{({\bm p}^2)^{\gamma -1 - \epsilon}} \left[\frac{1}{\epsilon} +\chi_0(\gamma) \right],  \notag \\
   \int \frac{d^{2 + 2 \epsilon}{\bm r}}{\pi^{1 + \epsilon}}
  \frac{\ln(1/{\bm r}^2)}{{\bm r}^2  [({\bm p} + {\bm r})^2]^{1-\gamma}}          
&=
\frac{\Gamma(1-\epsilon)}{({\bm p}^2)^{\gamma -1 - \epsilon}} 
\bigg[\frac{1}{\epsilon^2} 
-
 \frac{\ln {\bm p}^2}{\epsilon}
\notag \\
&
\hspace{1cm}
 - \left ( \ln(\bm {p}^2)\chi_0 (\gamma) + \frac{\chi_0^2(\gamma) + \chi_0'(\gamma)}{2} + \frac{\pi^2}{6}  \right) + \mathcal{O}(\epsilon)
\bigg],
\end{align}
with
\begin{align}
  \label{eq:chi0}
  \chi_0(\gamma) & =  2\psi(1) - \psi(\gamma) - \psi(1-\gamma),
\end{align}
the leading order BFKL characteristic function. We therefore find that
the proposed subtraction yields indeed a finite 
result.

\bibliographystyle{spphys} 
\bibliography{paper}

\begin{thebibliography}{10}
\providecommand{\url}[1]{{#1}}
\providecommand{\urlprefix}{URL }
\expandafter\ifx\csname urlstyle\endcsname\relax
  \providecommand{\doi}[1]{DOI \discretionary{}{}{}#1}\else
  \providecommand{\doi}{DOI \discretionary{}{}{}\begingroup
  \urlstyle{rm}\Url}\fi

\bibitem{Aad:2012tfa}
G.~Aad, et~al., Phys. Lett. B \textbf{716}, 1 (2012).
\newblock \doi{10.1016/j.physletb.2012.08.020}

\bibitem{Chatrchyan:2012ufa}
S.~Chatrchyan, et~al., Phys. Lett. B \textbf{716}, 30 (2012).
\newblock \doi{10.1016/j.physletb.2012.08.021}

\bibitem{DiMicco:2019ngk}
J.~Alison, et~al., in \emph{{Double Higgs Production at Colliders}}, ed. by
  B.~Di~Micco, M.~Gouzevitch, J.~Mazzitelli, C.~Vernieri (2019).
\newblock \doi{10.1016/j.revip.2020.100045}

\bibitem{Heinrich:2020ybq}
G.~Heinrich,   (2020)

\bibitem{Dreyer:2020xaj}
F.A. Dreyer, A.~Karlberg, J.N. Lang, M.~Pellen,   (2020)

\bibitem{Monni:2020nks}
P.F. Monni, E.~Re, M.~Wiesemann,   (2020)

\bibitem{Catani:1990eg}
S.~Catani, M.~Ciafaloni, F.~Hautmann, Nucl. Phys. \textbf{B366}, 135 (1991).
\newblock \doi{10.1016/0550-3213(91)90055-3}

\bibitem{Collins:1991ty}
J.C. Collins, R.K. Ellis, Nucl. Phys. \textbf{B360}, 3 (1991).
\newblock \doi{10.1016/0550-3213(91)90288-9}

\bibitem{Lipatov:1995pn}
L.N. Lipatov, Nucl. Phys. \textbf{B452}, 369 (1995).
\newblock \doi{10.1016/0550-3213(95)00390-E}

\bibitem{Lipatov:1996ts}
L.N. Lipatov, Phys. Rept. \textbf{286}, 131 (1997).
\newblock \doi{10.1016/S0370-1573(96)00045-2}

\bibitem{Celiberto:2020tmb}
F.G. Celiberto, D.Y. Ivanov, M.M. Mohammed, A.~Papa,   (2020)

\bibitem{Mangano:2017tke}
 \textbf{3/2017} (2017).
\newblock \doi{10.23731/CYRM-2017-003}

\bibitem{Fadin:1975cb}
V.S. Fadin, E.A. Kuraev, L.N. Lipatov, Phys. Lett. \textbf{60B}, 50 (1975).
\newblock \doi{10.1016/0370-2693(75)90524-9}

\bibitem{Lipatov:1976zz}
L.~Lipatov, Sov. J. Nucl. Phys. \textbf{23}, 338 (1976)

\bibitem{Kuraev:1977fs}
E.A. Kuraev, L.N. Lipatov, V.S. Fadin, Sov. Phys. JETP \textbf{45}, 199 (1977).
\newblock [Zh. Eksp. Teor. Fiz.72,377(1977)]

\bibitem{Balitsky:1978ic}
I.I. Balitsky, L.N. Lipatov, Sov. J. Nucl. Phys. \textbf{28}, 822 (1978).
\newblock [Yad. Fiz.28,1597(1978)]

\bibitem{Dumitru:2005gt}
A.~Dumitru, A.~Hayashigaki, J.~Jalilian-Marian, Nucl. Phys. \textbf{A765}, 464
  (2006).
\newblock \doi{10.1016/j.nuclphysa.2005.11.014}

\bibitem{Marquet:2007vb}
C.~Marquet, Nucl. Phys. A \textbf{796}, 41 (2007).
\newblock \doi{10.1016/j.nuclphysa.2007.09.001}

\bibitem{Deak:2009xt}
M.~Deak, F.~Hautmann, H.~Jung, K.~Kutak, JHEP \textbf{09}, 121 (2009).
\newblock \doi{10.1088/1126-6708/2009/09/121}

\bibitem{Chachamis:2015ona}
G.~Chachamis, M.~De\'ak, M.~Hentschinski, G.~Rodrigo, A.~Sabio~Vera, JHEP
  \textbf{09}, 123 (2015).
\newblock \doi{10.1007/JHEP09(2015)123}

\bibitem{Fadin:1998py}
V.S. Fadin, L.~Lipatov, Phys. Lett. B \textbf{429}, 127 (1998).
\newblock \doi{10.1016/S0370-2693(98)00473-0}

\bibitem{Ciafaloni:1998gs}
M.~Ciafaloni, G.~Camici, Phys. Lett. B \textbf{430}, 349 (1998).
\newblock \doi{10.1016/S0370-2693(98)00551-6}

\bibitem{Kovchegov:1999yj}
Y.V. Kovchegov, Phys. Rev. \textbf{D60}, 034008 (1999).
\newblock \doi{10.1103/PhysRevD.60.034008}

\bibitem{Balitsky:1995ub}
I.~Balitsky, Nucl. Phys. \textbf{B463}, 99 (1996).
\newblock \doi{10.1016/0550-3213(95)00638-9}

\bibitem{JalilianMarian:1997jx}
J.~Jalilian-Marian, A.~Kovner, A.~Leonidov, H.~Weigert, Nucl. Phys.
  \textbf{B504}, 415 (1997).
\newblock \doi{10.1016/S0550-3213(97)00440-9}

\bibitem{JalilianMarian:1997gr}
J.~Jalilian-Marian, A.~Kovner, A.~Leonidov, H.~Weigert, Phys. Rev.
  \textbf{D59}, 014014 (1998).
\newblock \doi{10.1103/PhysRevD.59.014014}

\bibitem{Kovner:2000pt}
A.~Kovner, J.G. Milhano, H.~Weigert, Phys. Rev. \textbf{D62}, 114005 (2000).
\newblock \doi{10.1103/PhysRevD.62.114005}

\bibitem{Kovner:1999bj}
A.~Kovner, J.G. Milhano, Phys. Rev. \textbf{D61}, 014012 (2000).
\newblock \doi{10.1103/PhysRevD.61.014012}

\bibitem{Ducloue:2013hia}
B.~Ducloue, L.~Szymanowski, S.~Wallon, JHEP \textbf{05}, 096 (2013).
\newblock \doi{10.1007/JHEP05(2013)096}

\bibitem{Caporale:2016zkc}
F.~Caporale, F.~Celiberto, G.~Chachamis, D.G. Gomez, A.~Sabio~Vera, Phys. Rev.
  D \textbf{95}(7), 074007 (2017).
\newblock \doi{10.1103/PhysRevD.95.074007}

\bibitem{Celiberto:2016ygs}
F.G. Celiberto, D.Y. Ivanov, B.~Murdaca, A.~Papa, Eur. Phys. J. C
  \textbf{76}(4), 224 (2016).
\newblock \doi{10.1140/epjc/s10052-016-4053-5}

\bibitem{Chachamis:2017vfa}
G.~Chachamis, F.~Caporale, F.G. Celiberto, D.~Gordo~Gomez, A.~Sabio~Vera, PoS
  \textbf{DIS2017}, 067 (2018).
\newblock \doi{10.22323/1.297.0067}

\bibitem{Caporale:2018qnm}
F.~Caporale, F.~Celiberto, G.~Chachamis, D.~Gordo~G\'omez, A.~Sabio~Vera, Nucl.
  Phys. B \textbf{935}, 412 (2018).
\newblock \doi{10.1016/j.nuclphysb.2018.09.002}

\bibitem{vanHameren:2014ala}
A.~van Hameren, P.~Kotko, K.~Kutak, S.~Sapeta, Phys. Lett. B \textbf{737}, 335
  (2014).
\newblock \doi{10.1016/j.physletb.2014.09.005}

\bibitem{vanHameren:2015uia}
A.~van Hameren, P.~Kotko, K.~Kutak, Phys. Rev. D \textbf{92}(5), 054007 (2015).
\newblock \doi{10.1103/PhysRevD.92.054007}

\bibitem{Bautista:2016xnp}
I.~Bautista, A.~Fernandez~Tellez, M.~Hentschinski, Phys. Rev. D \textbf{94}(5),
  054002 (2016).
\newblock \doi{10.1103/PhysRevD.94.054002}

\bibitem{Celiberto:2018muu}
F.~Celiberto, D.~Gordo~G\'omez, A.~Sabio~Vera, Phys. Lett. B \textbf{786}, 201
  (2018).
\newblock \doi{10.1016/j.physletb.2018.09.045}

\bibitem{Garcia:2019tne}
A.~Arroyo~Garcia, M.~Hentschinski, K.~Kutak, Phys. Lett. B \textbf{795}, 569
  (2019).
\newblock \doi{10.1016/j.physletb.2019.06.061}

\bibitem{Bartels:2001ge}
J.~Bartels, D.~Colferai, G.~Vacca, Eur. Phys. J. C \textbf{24}, 83 (2002).
\newblock \doi{10.1007/s100520200919}

\bibitem{Bartels:2002yj}
J.~Bartels, D.~Colferai, G.~Vacca, Eur. Phys. J. C \textbf{29}, 235 (2003).
\newblock \doi{10.1140/epjc/s2003-01169-5}

\bibitem{Hentschinski:2014esa}
M.~Hentschinski, J.D.M. Mart\'\i{}nez, B.~Murdaca, A.~Sabio~Vera, Nucl. Phys. B
  \textbf{889}, 549 (2014).
\newblock \doi{10.1016/j.nuclphysb.2014.10.026}

\bibitem{Hentschinski:2014bra}
M.~Hentschinski, J.~Madrigal~Mart\'\i{}nez, B.~Murdaca, A.~Sabio~Vera, Nucl.
  Phys. B \textbf{887}, 309 (2014).
\newblock \doi{10.1016/j.nuclphysb.2014.08.010}

\bibitem{Hentschinski:2014lma}
M.~Hentschinski, J.D. Madrigal~Martínez, B.~Murdaca, A.~Sabio~Vera, Phys.
  Lett. \textbf{B735}, 168 (2014).
\newblock \doi{10.1016/j.physletb.2014.06.022}

\bibitem{Chachamis:2012cc}
G.~Chachamis, M.~Hentschinski, J.D. Madrigal~Mart\'\i{}nez, A.~Sabio~Vera,
  Phys. Rev. D \textbf{87}(7), 076009 (2013).
\newblock \doi{10.1103/PhysRevD.87.076009}

\bibitem{Celiberto:2017ptm}
F.G. Celiberto, D.Y. Ivanov, B.~Murdaca, A.~Papa, Eur. Phys. J. C
  \textbf{77}(6), 382 (2017).
\newblock \doi{10.1140/epjc/s10052-017-4949-8}

\bibitem{Boussarie:2019ero}
R.~Boussarie, A.~Grabovsky, L.~Szymanowski, S.~Wallon, Phys. Rev. D
  \textbf{100}(7), 074020 (2019).
\newblock \doi{10.1103/PhysRevD.100.074020}

\bibitem{Boussarie:2016ogo}
R.~Boussarie, A.~Grabovsky, L.~Szymanowski, S.~Wallon, JHEP \textbf{11}, 149
  (2016).
\newblock \doi{10.1007/JHEP11(2016)149}

\bibitem{Beuf:2017bpd}
G.~Beuf, Phys. Rev. D \textbf{96}(7), 074033 (2017).
\newblock \doi{10.1103/PhysRevD.96.074033}

\bibitem{vanHameren:2012uj}
A.~van Hameren, P.~Kotko, K.~Kutak, JHEP \textbf{12}, 029 (2012).
\newblock \doi{10.1007/JHEP12(2012)029}

\bibitem{vanHameren:2012if}
A.~van Hameren, P.~Kotko, K.~Kutak, JHEP \textbf{01}, 078 (2013).
\newblock \doi{10.1007/JHEP01(2013)078}

\bibitem{vanHameren:2013csa}
A.~van Hameren, K.~Kutak, T.~Salwa, Phys. Lett. \textbf{B727}, 226 (2013).
\newblock \doi{10.1016/j.physletb.2013.10.039}

\bibitem{vanHameren:2016kkz}
A.~van Hameren, Comput. Phys. Commun. \textbf{224}, 371 (2018).
\newblock \doi{10.1016/j.cpc.2017.11.005}

\bibitem{Hentschinski:2020rfx}
M.~Hentschinski,   (2020)

\bibitem{Bock:2020cnd}
M.G. Bock, M.~Hentschinski, A.~Sabio~Vera,   (2020)

\bibitem{Hautmann:2002tu}
F.~Hautmann, Phys. Lett. B \textbf{535}, 159 (2002).
\newblock \doi{10.1016/S0370-2693(02)01761-6}

\bibitem{Lipatov:2005at}
A.~Lipatov, N.~Zotov, Eur. Phys. J. C \textbf{44}, 559 (2005).
\newblock \doi{10.1140/epjc/s2005-02393-7}

\bibitem{Lipatov:2009qx}
A.~Lipatov, N.~Zotov, Phys. Rev. D \textbf{80}, 013006 (2009).
\newblock \doi{10.1103/PhysRevD.80.013006}

\bibitem{Lipatov:2015vya}
A.~Lipatov, N.~Zotov, Eur. Phys. J. C \textbf{75}(5), 189 (2015).
\newblock \doi{10.1140/epjc/s10052-015-3419-4}

\bibitem{Chachamis:2012mw}
G.~Chachamis, M.~Hentschinski, J.D. Madrigal~Martínez, A.~Sabio~Vera, Phys.
  Part. Nucl. \textbf{45}(4), 788 (2014).
\newblock \doi{10.1134/S1063779614040030}

\bibitem{Hentschinski:2011tz}
M.~Hentschinski, A.~Sabio~Vera, Phys. Rev. \textbf{D85}, 056006 (2012).
\newblock \doi{10.1103/PhysRevD.85.056006}

\bibitem{Chachamis:2012gh}
G.~Chachamis, M.~Hentschinski, J.D. Madrigal~Martinez, A.~Sabio~Vera, Nucl.
  Phys. \textbf{B861}, 133 (2012).
\newblock \doi{10.1016/j.nuclphysb.2012.03.015}

\bibitem{Chachamis:2013hma}
G.~Chachamis, M.~Hentschinski, J.D. Madrigal~Martinez, A.~Sabio~Vera, Nucl.
  Phys. \textbf{B876}, 453 (2013).
\newblock \doi{10.1016/j.nuclphysb.2013.08.013}

\bibitem{Hentschinski:2011xg}
M.~Hentschinski, Nucl. Phys. \textbf{B859}, 129 (2012).
\newblock \doi{10.1016/j.nuclphysb.2012.02.001}

\bibitem{Lipatov:2000se}
L.N. Lipatov, M.I. Vyazovsky, Nucl. Phys. \textbf{B597}, 399 (2001).
\newblock \doi{10.1016/S0550-3213(00)00709-4}

\bibitem{Nefedov:2017qzc}
M.~Nefedov, V.~Saleev, Mod. Phys. Lett. \textbf{A32}(40), 1750207 (2017).
\newblock \doi{10.1142/S0217732317502078}

\bibitem{Nefedov:2018vyt}
M.~Nefedov, V.~Saleev, Phys. Lett. \textbf{B790}, 551 (2019).
\newblock \doi{10.1016/j.physletb.2018.12.071}

\bibitem{Nefedov:2019mrg}
M.A. Nefedov, Nucl. Phys. \textbf{B946}, 114715 (2019).
\newblock \doi{10.1016/j.nuclphysb.2019.114715}

\bibitem{Ellis:1975ap}
J.R. Ellis, M.K. Gaillard, D.V. Nanopoulos, Nucl. Phys. B \textbf{106}, 292
  (1976).
\newblock \doi{10.1016/0550-3213(76)90382-5}

\bibitem{Shifman:1979eb}
M.A. Shifman, A.~Vainshtein, M.~Voloshin, V.I. Zakharov, Sov. J. Nucl. Phys.
  \textbf{30}, 711 (1979)

\bibitem{Dawson:1990zj}
S.~Dawson, Nucl. Phys. B \textbf{359}, 283 (1991).
\newblock \doi{10.1016/0550-3213(91)90061-2}

\bibitem{Ravindran:2002dc}
V.~Ravindran, J.~Smith, W.~Van~Neerven, Nucl. Phys. B \textbf{634}, 247 (2002).
\newblock \doi{10.1016/S0550-3213(02)00333-4}

\bibitem{Catani:1994sq}
S.~Catani, F.~Hautmann, Nucl. Phys. B \textbf{427}, 475 (1994).
\newblock \doi{10.1016/0550-3213(94)90636-X}

\bibitem{Ciafaloni:1998hu}
M.~Ciafaloni, D.~Colferai, Nucl. Phys. B \textbf{538}, 187 (1999).
\newblock \doi{10.1016/S0550-3213(98)00621-X}

\bibitem{Catani:1996vz}
S.~Catani, M.~Seymour, Nucl. Phys. B \textbf{485}, 291 (1997).
\newblock \doi{10.1016/S0550-3213(96)00589-5}.
\newblock [Erratum: Nucl.Phys.B 510, 503--504 (1998)]

\bibitem{Fadin:1998fv}
V.S. Fadin, R.~Fiore, Phys. Lett. B \textbf{440}, 359 (1998).
\newblock \doi{10.1016/S0370-2693(98)01099-5}

\bibitem{Bartels:2006hg}
J.~Bartels, A.~Sabio~Vera, F.~Schwennsen, JHEP \textbf{11}, 051 (2006).
\newblock \doi{10.1088/1126-6708/2006/11/051}

\bibitem{Sun:2011iw}
P.~Sun, B.W. Xiao, F.~Yuan, Phys. Rev. D \textbf{84}, 094005 (2011).
\newblock \doi{10.1103/PhysRevD.84.094005}

\bibitem{Mueller:2013wwa}
A.~Mueller, B.W. Xiao, F.~Yuan, Phys. Rev. D \textbf{88}(11), 114010 (2013).
\newblock \doi{10.1103/PhysRevD.88.114010}

\bibitem{Xiao:2018esv}
B.W. Xiao, F.~Yuan, Phys. Lett. B \textbf{782}, 28 (2018).
\newblock \doi{10.1016/j.physletb.2018.04.070}

\bibitem{Nefedov:2020ecb}
M.~Nefedov, JHEP \textbf{08}, 055 (2020).
\newblock \doi{10.1007/JHEP08(2020)055}

\bibitem{Gituliar:2015agu}
O.~Gituliar, M.~Hentschinski, K.~Kutak, JHEP \textbf{01}, 181 (2016).
\newblock \doi{10.1007/JHEP01(2016)181}

\bibitem{Hentschinski:2016wya}
M.~Hentschinski, A.~Kusina, K.~Kutak, Phys. Rev. D \textbf{94}(11), 114013
  (2016).
\newblock \doi{10.1103/PhysRevD.94.114013}

\bibitem{Hentschinski:2017ayz}
M.~Hentschinski, A.~Kusina, K.~Kutak, M.~Serino, Eur. Phys. J. \textbf{C78}(3),
  174 (2018).
\newblock \doi{10.1140/epjc/s10052-018-5634-2}

\end{thebibliography}

\end{document}